\documentclass[fleqn,usenatbib]{mnras}

\usepackage[T1]{fontenc}

\usepackage{graphicx}	
\usepackage{amsmath}	
\usepackage{amssymb}	
\usepackage{tabularx}
\usepackage{booktabs}
\usepackage{multirow}
\usepackage{newtxtext,newtxmath}
\usepackage{comment}
\usepackage{soul}
\usepackage[dvipsnames]{xcolor}
\usepackage{enumitem}

\title[GRB afterglows with decaying microturbulence]{Numerical study of synchrotron and inverse-Compton radiation from gamma-ray burst afterglows with decaying microturbulence}

\author[Huang and Li]{
Yan Huang$^{1}$\thanks{E-mail: hyan623@ahu.edu.cn}
and Zhuo Li$^{2,3}$\thanks{E-mail: zhuo.li@pku.edu.cn}
\\
$^{1}$ School of Physics and Optoelectronics Engineering, Anhui University, Hefei 230601, China\\
$^{2}$ Department of Astronomy, School of Physics, Peking University, Beijing 100871, China\\
$^{3}$ Kavli Institute for Astronomy and Astrophysics, Peking University, Beijing 100871, China
}

\date{Accepted XXX. Received YYY; in original form ZZZ}

\pubyear{2022}

\begin{document}
\label{firstpage}
\pagerange{\pageref{firstpage}--\pageref{lastpage}}
\maketitle

\begin{abstract}


The multiwavelength observations of GRB afterglows, together with some high-performance particle-in-cell simulations, hint that the magnetic field may decay behind the shock front. In this work, we develop a numerical code to calculate the evolution of the accelerated electron distribution, their synchrotron and inverse-Compton (IC) spectra and accordingly the light curves (LCs) under the assumption of decaying microturbulence (DM) downstream of the shock, $\epsilon_B(t_p')\propto t_p'^{\alpha_t}$ with $t_p'$ the fluid proper time since injection. We find: 
(1) The synchrotron spectrum in the DM model is similar to that in the homogeneous turbulence (HT) model with very low magnetic field strength. However, the difference in the IC spectral component is relatively more obvious between them, due to the significant change of the postshock electron energy distribution with DM. 
(2) If the magnetic field decay faster, there are less electrons cool fast, and the IC spectral component becomes weaker. 
(3) The LCs in the DM model decay steeper than in the HT model, and the spectral evolution and the LCs in the DM model is similar to the HT model where the magnetic field energy fraction decreases with observer time, $\epsilon_B(t) \propto t^{5\alpha_t /8}$. 
(4) The DM model can naturally produce a significant IC spectral component in TeV energy range, but due to the Klein-Nishina suppression the IC power cannot be far larger than the synchrotron power. 
We apply the DM model to describe the afterglow data of GRB 190114C and find the magnetic field decay exponent $\alpha_t\sim -0.4$ and the electron spectral index $p\sim2.4$. 
Future TeV observations of the IC emission from GRB afterglows will further help to probe the poorly known microphysics of relativistic shocks.

\end{abstract}

\begin{keywords}
shock waves--gamma-ray bursts--radiation mechanisms: non-thermal
\end{keywords}


\section{Introduction}

Gamma-ray burst (GRB) afterglow emission is produced by a relativistic shock wave associated with the initial GRB explosion. A standard model for GRB afterglows has been well established \citep[see, e.g.,][for recent reviews]{kumar2015, Piran05}, in which the shock sweeps up the circum-burst material (CBM) and decelerates, whereas the electrons are accelerated to become relativistic and then radiate via synchrotron \citep[e.g.,][]{sari98, pan00} and inverse-Compton (IC) radiation \citep[e.g.,][]{sari01, nakar2009}, producing the GRB afterglow emission. Due to such a picture, GRB afterglows provide a laboratory to study the poorly known microphysics of relativistic collisionless shocks.

In the simple case that a spherical shock propagates into a homogeneous medium, the standard model can determine the GRB afterglow spectra given five physical parameters relevant to the shock, i.e., the total energy of the shock $E$, the CBM particle number density $n$, the energy fraction of the nonthermal electrons to the shock energy $\epsilon_{e}$, the energy fraction of the magnetic field to the shock energy $\epsilon_{B}$, and the power-law index of the newly accelerated electron distribution (ED) $p$ \citep[e.g.,][]{sari98, sari01}. In the simple analytical approach, the postshock ED is approximated as a broken-power-law, then the spectral energy distribution (SED) and the LCs also can be analytically expressed as broken power laws, which have been widely used to fit the observed data. The standard afterglow model provides a generally successful description of the observed data, thus it is widely believed that the afterglow emission is produced by the accelerated electrons in the GRB shock.

It is indeed with uncertainty about the magnetic field structure of the relativistic shocks. In the standard model of GRB afterglows, the magnetic field is assumed to be homogeneous downstream of the shock, and is expressed in terms of the equipartition parameter for postshock magnetic field,
\begin{equation}
\epsilon_{B}=\frac{B'^2}{32 \pi n m_{p} c^2 \Gamma^2},
\label{B}
\end{equation}
where $B'$ is the magnetic field strength downstream of the GRB shock (hereafter, the superscript prime ($'$) is used to denote the quantities in the rest frame of shocked fluid), and $\Gamma$ is the  bulk Lorentz factor of the shocked fluid. If the shock compression is the only cause to enhance the downstream magnetic field, then the magnetic field can be $B'=4 \Gamma B_{0}$ \citep{ach01}, where $B_{0}$ is the seed magnetic field in the CBM. For typical CBM with $B_0\sim$ a few $\rm \mu G$, and $n\sim1 \rm cm^{-3}$, one obtains $\epsilon_{B}\sim 10^{-9}$ for shock compression. However, the modeling of the multi-wavelength data of GRB afterglows usually result in much larger values, in the range of $\epsilon_B\sim 10^{-5} - 10^{-1}$ \citep[e.g.,][]{wij99,pan02,pan05,Santana14}, suggesting additional amplification of the postshock magnetic field. Except shock compression, there have been several theoretical and numerical studies that considered possible mechanisms operating in the plasma surrounding the GRB , and generating extra amplification of the magnetic field, e.g., the magnetic field of internal shock may originate from the central engine, but for the external shock, the magnetic field of the central engine is almost dissipated through the magnetic reconnection in the prompt emission stage, and cannot originate from the center. In addition, the magnetic fields of external shock can be amplified by Weibel instability\citep{wei59, med99, gru99, med05} and turbulence-generated dynamo \citep{sir07, goo08,zha09}.

The magnetic field is amplified in a small scales behind the shock front and then decayed rapidly through Landau damping\citep{chang08, lem2015, lemoine15}. Some observations and simulations support the fact that the magnetic field decays downstream of the shock front. Firstly, when compare the acceleration time to the synchrotron cooling time, we got the maximum energy of synchrotron photons at around 100s can only reach the order of GeV (since $E_{\rm max} \sim 100 \rm MeV \times \Gamma_b(t)/(1+z)$, $\Gamma_b$ is the Lorentz factor of GRB outflow, for the observed times around 100s, $\Gamma_b$ is order of 10). However, some long-lived ($\sim 100 -1000 \rm s$) GeV emissions of GRBs which detected by Fermi-LAT are originated from IC process, and the corresponding magnetic field strength $\epsilon_B$ is low, of the order of $10^{-6} -10^{-5}$ \citep{kumar2009, kumar2010, Barniol2011, he2011, liu2011}, much smaller than the canonical value $\epsilon_B \sim 0.01$. More direct evidence is the results of high-performance particle-in-cell (PIC) simulations, the PIC simulations show that the density in the downstream of the shock front is almost 4 times larger than the downstream, and the magnetic field strength $\epsilon_B$ is decayed with distance from the shock front \citep{spi08, Keshet09, mar09, sir09, sir11}, but the timescale and spacescale in the simulations are limited.

Since the electrons with different Lorentz factors cool at different time scales, corresponding to the regions of different magnetic field strengths, then this time dependence of $\epsilon_B$ will affect the spectra of GRB afterglows \citep{rossi2003, derishev2007, lemoine13b,lemoine13a, lemoine15}. \citet{lemoine13b, lemoine15} had carried out studies on the DM effects on the radiation signature of GRB afterglows. It is showed that in the DM model the electrons with different Lorentz factors can cool in regions with different magnetic fields, with the high (or low) energy photons being emitted by electrons that are close to (or far away from) the shock front, so the DM model shows different characteristic frequencies, temporal evolution and spectral slope compared with the standard model. Moreover, \citet{lemoine13a} analytically analyzed the multiwavelength LCs of four GRBs, detected with long-lived ($\sim \rm 100-1000s$) GeV emission, using approximation for synchrotron radiation in the power-law-decaying microturbulence ($\epsilon_{B} \propto t'^{\alpha_{t}}_p$, with $t_p'$ the proper time of the injected fluid), and determined that the decay power-law exponent is $-0.5 \lesssim \alpha_{t} \lesssim -0.4$.

In order to investigate in details the GRB afterglows with DM model, in this paper we carry out numerical calculation instead of the analytical approach.  We develop a numerical code to calculate the evolution of the EDs and SEDs, and the LCs of GRB afterglows in the DM model with a wide range of parameters, considering synchrotron and IC radiation with Klein-Nishina (KN) effect. We then compare the results of DM model with the observed data to constrain the magnetic field structure. 

The structures of this paper are as following. In section \ref{sec:review}, we review the basic physical model of GRB afterglows, including the hydrodynamics of GRB shock, the structure of DM and the radiative processes of GRB afterglows. In section \ref{sec:numerical}, we describe the numerical method and verify the correctness of our algorithm and code. In section \ref{results}, we perform numerical calculations by taking a wide range of testing parameters, comparing the DM and HT model, investigate the effects of magnetic field structure. In section \ref{apply}, we apply the DM model to GRB 190114C, which has the long-lived extented Sub-TeV emission, to constrain the magnetic field structure in the DM model. The conclusions are made in section \ref{conclusions}.

\section{Physical model of GRB afterglow} \label{sec:review}
\subsection{Hydrodynamic evolution of GRB shock}\label{shock evolution}

We assume a spherically symmetric system, which is a good assumption when the GRB fluid Lorentz factor and the jet opening angle satisfy $\Gamma>1/\theta_j$ \citep[e.g.,][]{rhoads99}. 
A relativistic ejecta is released by the GRB central engine and propagates into the CBM. The interaction of the ejecta with CBM generates a shock sweeping up the medium material. Denote the total ejecta energy as $E$ and the initial Lorentz factor as $\Gamma_0$. Consider the CBM as homogeneous medium with a constant proton number density $n$, typical for the interstellar medium (ISM). The shock transfers the ejecta energy into the swept up material. Consider the adiabatic shock case, i.e., the emission is negligible, the shock energy is \citep{bla76}
\begin{align}
E_{\rm sh}=\frac{16}{17} \pi \Gamma^2 R^3 n m_p c^2,
\label{E_n}
\end{align}
where $R$ is the shock radius, and $\Gamma$ is the post-shock material Lorentz factor.

Initially the shock propagates in constant speed, with $\Gamma\approx\Gamma_0$ and $E_{\rm sh}<E$. At the point that the shock sweeps up enough medium material and transfer most energy, $E_{\rm sh}\approx E$, the shock starts to decelerate significantly. The deceleration radius is
$
R_{\rm dec}=(17E/16\pi \Gamma_0^2 n m_p c^2)^{1/3}.
$
Using the relation $R=4 c \Gamma^2 t_s$, which takes into account the effect of the equal arrival time surface \citep{wax1997}, with $t_s$ being the observer's time measured by an observer in rest with the GRB source, the deceleration time is given by
\begin{align}
t_{\rm dec}=\left(\frac{17E}{1024 \pi n m_{p} c^5 \Gamma_0^8}\right)^{1/3},
\label{t_dec}
\end{align}
after which the shock dynamics transits into self-similar solution \citep{bla76}. To summarize, we consider the simple shock dynamical evolution as follows,
\begin{align}
R \approx
\begin{cases}
4 c \Gamma_{0}^2 t_s, & t_s \leq t_{\rm dec}\\
\left(\frac{17E}{4 \pi m_{p} n c}\right)^{1/4}t_s^{1/4}, & t_s>t_{\rm dec}
\end{cases}
\label{R_dyn}
\end{align}

\begin{equation}
\Gamma \approx
\begin{cases}
\Gamma_{0}, & t_s \leq t_{\rm dec}\\
\left(\frac{17E}{1024 \pi n m_{p} c^5}\right)^{1/8}t_s^{-3/8}, & t_s>t_{\rm dec}
\end{cases}
\label{dynamic}
\end{equation}

The observer's time $t$ and the Lorentz factor of the shocked fluid are related to the fluid  comoving time $t'$ according to
\begin{align}
dt_s=\Gamma (1-\beta) dt' \approx \frac1{2\Gamma} dt',
\label{t_co}
\end{align}
where $\beta=\sqrt{1-1/\Gamma}$ is the dimensionless velocity of the shock. As for the observer on the Earth, the observer's time ($t$) should be $dt=dt_s(1+z)$ for GRBs with cosmological redshift $z$.

\subsection{Magnetic field structure}\label{magnetic field}

In the early works, a standard GRB afterglow model is set up, where a homogeneous magnetic field is assumed behind the shock front and $\epsilon_{B}  \sim 10^{-2}$ near the shock front. But some theoretical works suggest that through Landau damping, the microturbulence behind the shock front may decay as a power-law\citep{chang08,lemoine13b,lemoine15}, i.e., $\epsilon_B \simeq \epsilon_{B+}(t'_{\rm p}/t'_{\rm \mu +})^{\alpha_t}$, where $t_{\rm p}'$ is the proper time since the electrons are injected into the shock in the rest frame of the shocked fluid. 
We here follow \cite{lemoine13b} to assume that the postshock fluid flows away downstream with a convective velocity $c/3$ relative to the shock front, and describe the magnetic field strength $B'$ in the rest frame of the fluid as a power-law decay, i.e.,
\begin{equation}
B' =
\begin{cases}
B'_{\mu} & \mbox{$t_{\rm p}' \leq t'_{\rm \mu +}$}\\
B'_{\mu} \left(t_{\rm p}' / t'_{\rm \mu +}\right)^{\alpha_{t}/2} & \mbox{$t'_{\rm \mu +}< t_{\rm p}' < t'_{\rm \mu -}$}\\
B'_{d} & \mbox{$t_{\rm p}' \geq t'_{\rm \mu -}$},
\end{cases}
\label{B}
\end{equation}
Here $B_{\mu}'$ is the magnetic field strength immediately behind the shock front, for which we define the equipartition parameter $\epsilon_{B+}$,
\begin{equation}
B'_{\mu}=(32 \pi m_{p} \epsilon_{B+} n)^{1/2} \Gamma c.
\label{Bup}
\end{equation}
Crossing the shock front, the magnetic field strength keeps a constant value $\epsilon_B=\epsilon_{B+}=10^{-2}$ due to the turbulence near the shock front until $t'_{\rm \mu+}$. We define $t'_{\rm \mu +} = \Delta_{\mu} \omega_{\rm pi}^{-1}$, where $\omega_{\rm pi}\approx(4 \pi n e^2/m_p)^{1/2}$ is the postshock plasma frequency. 
According to PIC simulations, the characteristic scale $\Delta_{\mu}$ separating far from the shock front is $\Delta_{\mu} \sim 10^2- 10^3$ \citep{chang08,Keshet09,lemoine13b}. Moreover, $t'_{\rm \mu-}$ is the proper time at which the field has relaxed to the background shock-compressed value $B'(t'_{\mu-})=B'_d=4 \Gamma B_{0}$, where $B_0$ is the upstream magnetic field strength, and $B_{0}=\rm 10^{-5} G$ is taken in this work.

PIC simulations in \citet{chang08} and \citet{Keshet09} suggest that $-1< \alpha_{t} <0$. However, given the present limitations of the PIC simulations, and the possible caveat related to the extension of the magnetic perturbation spectrum, one cannot exclude yet $\alpha_{t}<-1$. And some simulations of the development and the dynamics of relativistic Weibel turbulence indeed suggest a value $\alpha_{t} \approx -2$ \citep{Medvedev11}. In order to account for a broad possibilities in the DM model, we will consider both cases of $\alpha_{t}<-1$ and $-1<\alpha_{t}<0$ in the following discussion.

\subsection{Energy distribution of injected electrons}

According to the Fermi acceleration mechanism of relativistic collisionless shocks, the swept-up charged particles are expected to be accelerated to a power-law distribution in energy, and the acceleration zone is with thickness of about $\sim 100 c/ \omega_{\rm pi}$ \citep{lemoine13b, lemoine15}. We assume the energy distribution of electrons injected into the shock follow a power-law between $\gamma_m'$ and $\gamma_{\rm max}'$($\gamma_{\rm max}'>\gamma_{m}'$),
\begin{align}
d \dot N_{\rm e,0}=\dot N_{\rm e} \frac{p-1}{\gamma_{\rm m}'} \left(\frac{\gamma_{e}'}{\gamma_{\rm m}'}\right)^{-p} d \gamma_{e}',
\end{align}
with
\begin{align}
\dot N_{e}=\Gamma \left(\beta +\frac{1}{3}\right)4 \pi n R^2 c
\end{align}
being the number of electrons swept up and accelerated by the shock wave per unit time, as measured in the comoving downstream frame, $\gamma_{e}'$ is the random Lorentz factor of an electron, $\gamma_{m}'$ is the minimum Lorentz factor which can be determined by the shock jump conditions\citep{bla76, sari98},
\begin{align}
\gamma_{m}'=\frac{p-2}{p-1} \frac{m_{p}}{m_{e}}\epsilon_{e} \Gamma ,
\label{gamma_m}
\end{align}
and $\gamma_{\rm max}'$ is the maximum Lorentz factor that can be accelerated to. The acceleration time scale can be estimated by a correction factor $k_B$ multiplying the electron's Lamor timescale in the downstream region, $t_{\rm acc}' \approx k_B \gamma'_{e} m_e c/e B'_{\mu}$, where $k_B\gtrsim 1$.
The electron acceleration is mainly constrained by the electron's radiative energy loss. Due to strong KN suppression of IC scatterings at these high energies, it is reasonable to assume that the synchrotron cooling dominates around $\gamma_{\rm max}'$. The synchrotron cooling timescale is $t_{\rm syn}'=6 \pi m_{e} c/\gamma'_{e} \sigma_{\rm T} B_{\mu}'^2$. By $t_{\rm acc}'=t_{\rm syn}'$ we have
\begin{align}
\gamma_{\rm max}'=\sqrt{\frac{6 \pi e}{k_B \sigma_T B'_{\mu}}}.
\label{gamma_max}
\end{align}


\subsection{Temporal evolution of electron distribution}\label{sec:ED}

The accelerated electrons are continuously injected into the downstream of the shock and produce emission when flowing away from the shock front. In the standard GRB afterglow model where the downstream magnetic field is homogeneous, the numerical calculations for the afterglow emission usually apply the one-zone time-dependent model \citep[e.g.,][]{pet09, pen14, fuk17}. However in the DM model, the magnetic field strength, or $\epsilon_B$, decays with the distance away from the shock front, thus the one-zone approximation is invalid. The treatment should account for the fact that electrons of different energy may mainly cool and radiate at different timescales hence in regions with different magnetic field strengths. 

Here, we decompose the electrons into a series of thin shells that are injected into the downstream of the shock subsequently. We calculate the temporal evolution of the ED for each shell individually. For a certain shell, once injected into the shock downstream, the temporal evolution of the ED $dN_{e}/d\gamma_{e}'$ is governed by the continuity equation in the energy space,


\begin{equation}
\frac{\partial}{\partial t'} \frac{dN_{e}}{d\gamma_{e}'} +\frac{\partial}{\partial \gamma_{\rm e }'} \left(\dot \gamma_{\rm e}' \frac{dN_{e}}{d\gamma_{e}'}  \right) =0,
\label{continuity equation}
\end{equation}
where $t'$ is the time measured in the rest frame of the downstream fluid, and $\dot \gamma_{\rm e}'$ is the varying rate of an electron's Lorentz factor. Processes that lead to electron cooling include the adiabatic cooling ($\dot \gamma_{\rm e,adi}'$), synchrotron cooling ($\dot \gamma_{\rm e,syn}'$) and IC cooling ($\dot \gamma_{\rm e, ic}$), which will be discussed later. We will neglect any electron heating, e.g., due to synchrotron self-absorption. Thus we have
\begin{align}
\dot \gamma_{\rm e}'= \dot \gamma_{\rm e,adi}' +\dot \gamma_{\rm e,syn}'+\dot \gamma_{\rm e, ic}'.
\label{gamma_dot}
\end{align}
Once injected the electron number of a shell is conserved, because the electron-positron pair production due to $\gamma \gamma$-absorption is negligible and not considered, and there is no more injection, thus the source term in the right hand side of Eq. (\ref{continuity equation}) is zero.

\subsection{Cooling and radiation} \label{sec:SED}

\subsubsection{Adiabatic cooling}

The adiabatic cooling rate due to the spreading of the fluid is given by $\dot{\gamma}'_{\rm e, adi}=(1/3) \gamma'_{e} d \ln n'_e/dt'$. For a relativistic shock propagating in a homogeneous CBM with medium density $n$, the postshock electron number density of the downstream fluid ($n'_e$; in the comoving frame of the postshock fluid) is $n'_e \simeq 4 \Gamma n$ \citep{bla76}. For simplicity, assume the density of the postshock fluid is constant independent of the distance from the shock front. According to the hydrodynamic evolution of the relativistic shock, $\Gamma \propto R^{-3/2}$, we obtain $n'_e \propto R^{-3/2}$, and then $\dot{\gamma}'_{\rm e, adi}=-(1/2)(\gamma'_e/R)dR/dt'$. With $dR/dt'=\beta c\Gamma$, we have
\begin{equation}
\dot{\gamma}'_{\rm e, adi}=- \frac{\beta c\gamma'_e  \Gamma}{2R} .
\label{adi}
\end{equation}

\subsubsection{Synchrotron cooling and Synchrotron radiation}

The relativistic shock can compress and amplify the CBM magnetic field, in which the relativistic electrons give rise to synchrotron radiation. The synchrotron cooling rate of an electron with Lorentz factor $\gamma_{e}'$ in a magnetic field $B'$ is
\begin{equation}
\dot \gamma'_{\rm e, syn}=- \frac{\sigma_{T} B'^2 \gamma_{e}'^2}{6 \pi m_{e} c}.
\label{syn}
\end{equation}
In the comoving frame of the fluid, the shell with ED $dN_{e} / d\gamma_{e}'$ produces synchrotron radiation with the emitted power given by the integral of the individual spectral power per electron $\mathcal{R} (\nu'/\nu_{c}')$ over the ED of the electrons. The synchroton power per unit frequency at frequency $\nu'$ by the shell is written as \citep{ryb79}
\begin{equation}
\delta P_{\rm syn}'(\nu')=\frac{\sqrt{3} e^3 B'}{m_{e} c^2} \int d\gamma_e' \frac{dN_{e}}{d\gamma_{e}'} \mathcal{R}  \left(\frac{\nu'}{\nu_{c}'}\right),
\label{P_syn}
\end{equation}
where $\nu_c'=3eB' \gamma_e'^2/ 4\pi m_{e} c$ is the critical frequency. \citet{cru86} derives an exact expression for $\mathcal{R} (\nu'/\nu_{c}')$ in terms of Whittaker's function, which can be presented in a simple analytical approximation \citep{zir07},
\begin{equation}
\mathcal{R} \left(\frac{\nu'}{\nu_{c}'}\right)=\frac{1.81\exp (-\nu'/\nu_{c}')}{\sqrt{ (\nu'/\nu_{c}')^{-2/3}+(3.62/\pi)^2}}.
\end{equation}

\subsubsection{IC cooling and IC radiation}\label{lc cooling}

The electron cooling rate due to IC scatterings off a seed photon field with photon spectrum $n_{\nu'}$ is, considering only the first-order IC component and the KN correction in high energy, 
\begin{align}
\dot{\gamma}'_{\rm e, ic}=-\frac{1}{m_{e}c^2} \frac{3 \sigma_{T} c}{4 \gamma_{e}'^2} \int^{\nu_{\rm max}'}_{\nu_{\rm min}'} \frac{n_{\nu'} d\nu'}{\nu'} \int^{\nu_{\rm ic, max}'}_{\nu_{\rm ic, min}'} h \nu_{\rm ic}' d \nu_{\rm ic}' F(q,g),
\label{ssc}
\end{align}
where $\nu'$ and $\nu'_{\rm ic}$ are the frequencies of the seed photons and the IC scattered photons, respectively \citep{blu70, fan08,geng18}, and $n_{\nu'}$ is the seed photon density per unit frequency that the electrons encounter. For KN correction, we take
\begin{equation}
F(q,g)=2q \ln q+(1+2q)(1-q)+\frac{1}{2} \frac{(4qg)^2}{1+4gq}(1-q),
\end{equation}
with $g=\gamma_{e}' h \nu'/m_{e} c^2$, $w=h \nu_{\rm ic}'/\gamma_{e}' m_{e} c^2$, and $q=w/4g(1-w)$.  According to the dynamics of the collision between a relativistic electron and a photon, i.e., $1 \ll h \nu'/ \gamma_{e}' m_{e} c^2 \leq h \nu_{\rm ic}'/ \gamma_{e}' m_{e} c^2 \leq 4g/(1+4g)$, and $1/4 \gamma_{e}' \leq q \leq 1$, the upper limit of the second integral can be derived to be $h \nu_{\rm ic,max}'=\gamma_{e}' m_{e} c^2 (4g/4g+1)$, and the lower limit be $\nu_{\rm ic,min}'=\nu'$.

The seed photons for IC scatterings are the synchrotron photons contributed from all already injected shells. For a certain shell with ED $dN_e/d\gamma_e'$, the specific synchrotron power, $\delta P'_{\rm syn}(\nu')$, can be given by eq. (\ref{P_syn}). If $W'$ is the shell's thickness in the rest frame, the time that the photons stay in the shell is $W'/c$, thus the photon density can be estimated by the total photon number produced during the staying time divided by the shell's volume $4 \pi  R^2W'$. The specific number density of the synchrotron photons contributed by this certain shell is then
\begin{align}
\delta n_{\nu'} \approx \frac{\delta P'_{\rm syn}(\nu')}{h \nu'4\pi R^2c},
\label{n_nv}
\end{align}
where $W'$ is canceled out. The total seed density $n_{\nu'}$ in eq. (\ref{ssc}) should be the sum of all the relevant shells,  $n_{\nu'}=\int\delta n_{\nu'}$. 

The scattered photon spectrum per electron can be expressed in terms of the seed photon spectrum $n_{\rm \nu'}$ by \citep{blu70}
\begin{align}
\frac{dN_{\gamma}'}{dt'd\nu_{\rm ic}'}=\frac{3\sigma_{T}cn_{\rm \nu'}}{4\gamma_{e}'^2\nu'} F(q,g)d\nu'.
\label{spec_per}
\end{align}
For a shell with the ED $dN_{e} / d\gamma_{e}'$, the total SED of IC radiation is \citep{ryb79}
\begin{align}
\delta P'_{\rm ic}(\nu_{\rm ic}')=\int_{\nu'_{\rm min}}^{\nu'_{\rm max}} \int_{\gamma_{\rm e,min}'}^{\gamma_{\rm e,max}'}h \nu_{\rm ic}'  \frac{dN_{\gamma}'}  {dt'd\nu_{\rm ic}'d\nu'} \frac{dN_{\rm e}}{d\gamma_{e}'} d\gamma_{e}'d\nu'.
\label{P_ssc}
\end{align}

For a single shell, the total radiation power by all electrons in the shell is given by $\delta P'( \nu')=\delta P'_{\rm syn} (\nu')+\delta P'_{\rm ic} (\nu')$, where the synchrotron and IC power is given by Eq.(\ref{P_syn}) and Eq.(\ref{P_ssc}). The observed flux from the shell at frequency $\nu$ will be, ignoring the effect of the equal-arrival-time surface \citep{wax1997, gra1999},
\begin{equation}
\delta F_{ \nu} =\frac{(1+z) \delta P'(\nu) \Gamma }{4 \pi D_L^2},
\label{obs frame}
\end{equation}
with the observed photon frequency in relation with that in the rest frame of the fluid as
\begin{equation}
\nu =\frac{4 \Gamma}{3(1+z)} \nu',
\end{equation}
and $D_{L}$ being the luminosity distance\footnote{We adopt a flat $\Lambda$CDM universe, with $H_0=71~\rm km s^{-1}$, $\Omega_{m}=0.27$, and $\Omega_{\Lambda}=0.73$.}. The total observed flux is the sum of all shell's contribution, $F_\nu=\int\delta F_\nu$. 

\section{Numerical treatment} \label{sec:numerical}

\begin{figure*}
\vskip -0.0 true cm
\centering
\includegraphics[scale=0.4]{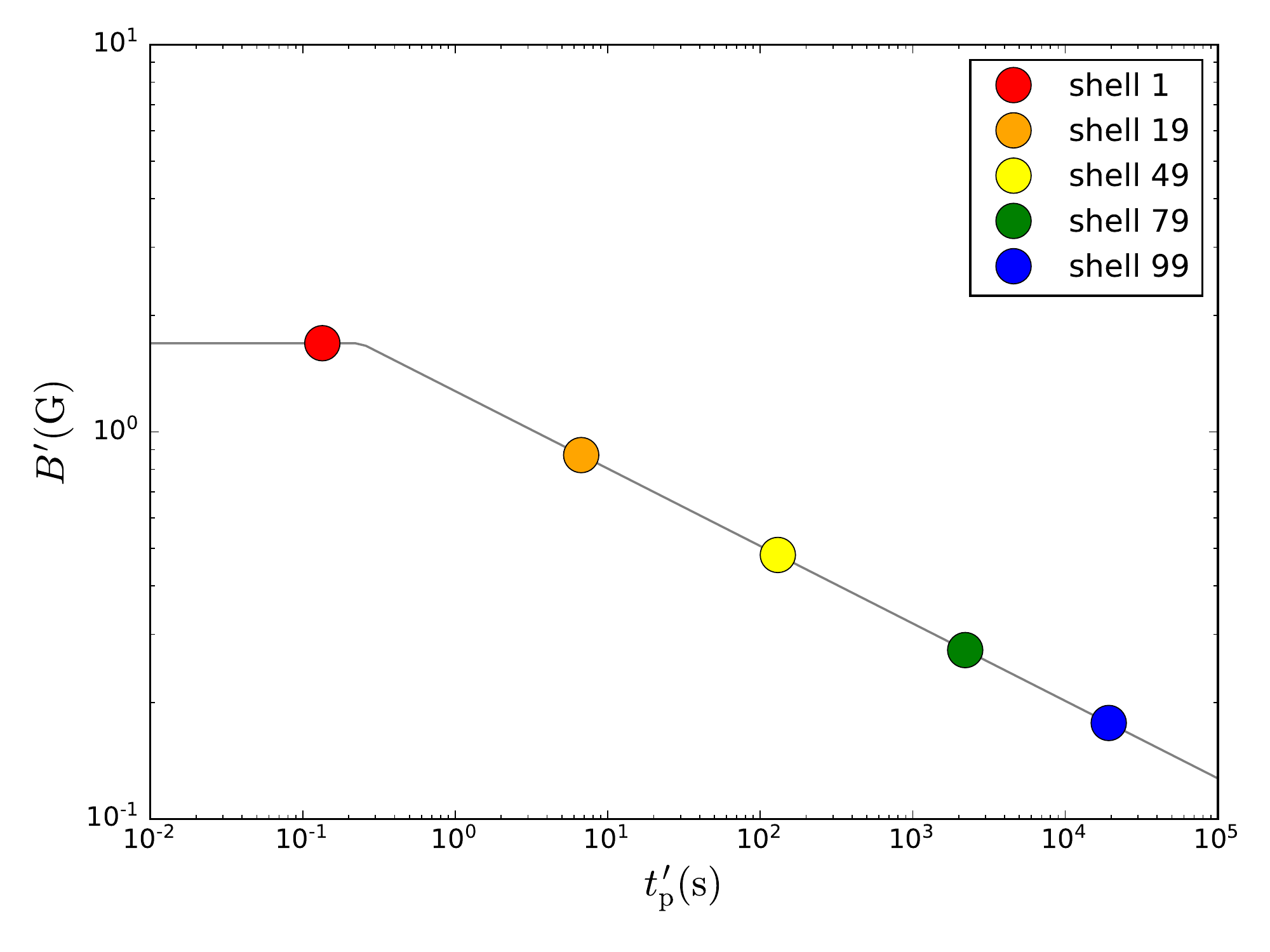} 
\includegraphics[scale=0.4]{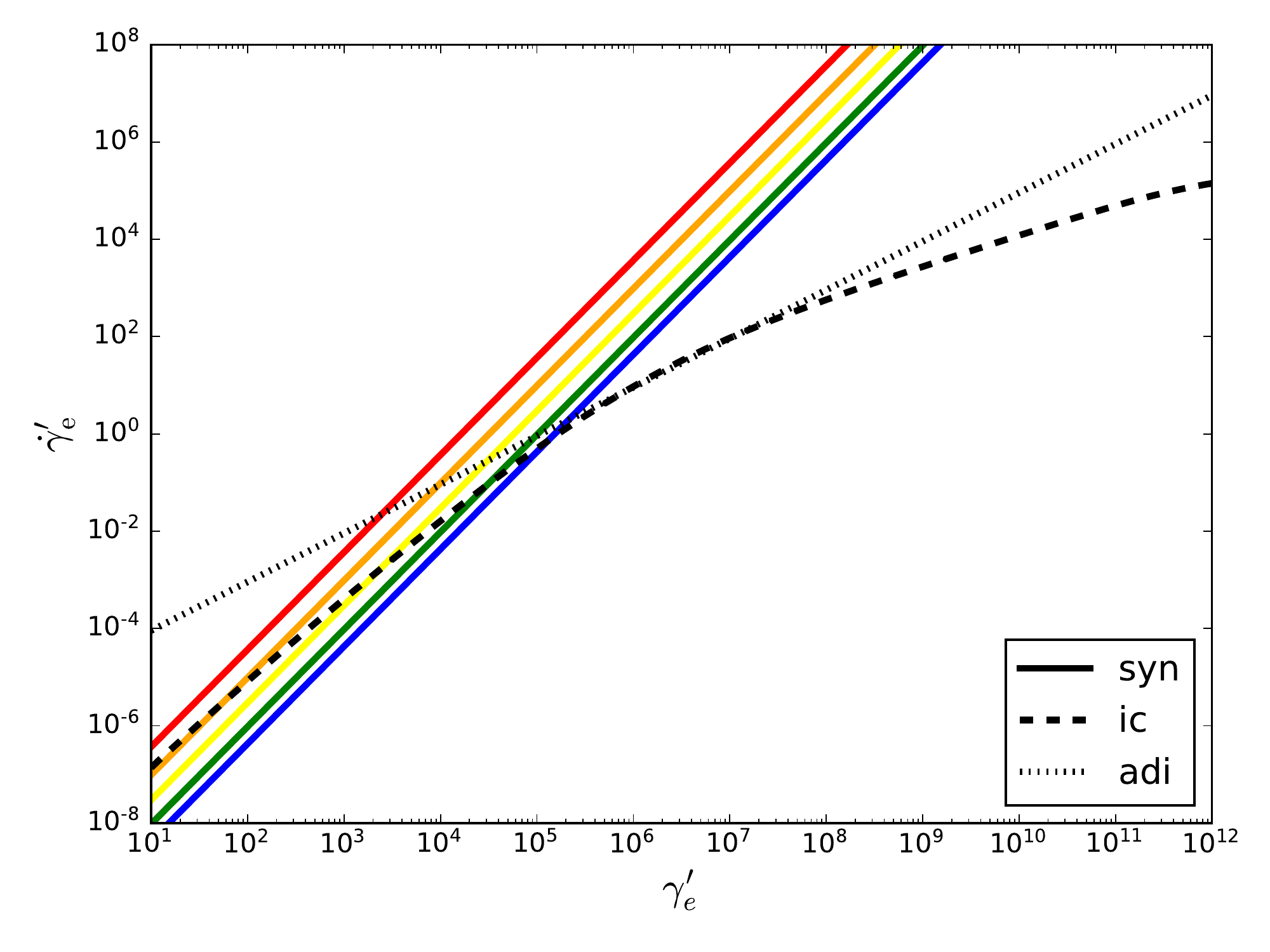}
\includegraphics[scale=0.4]{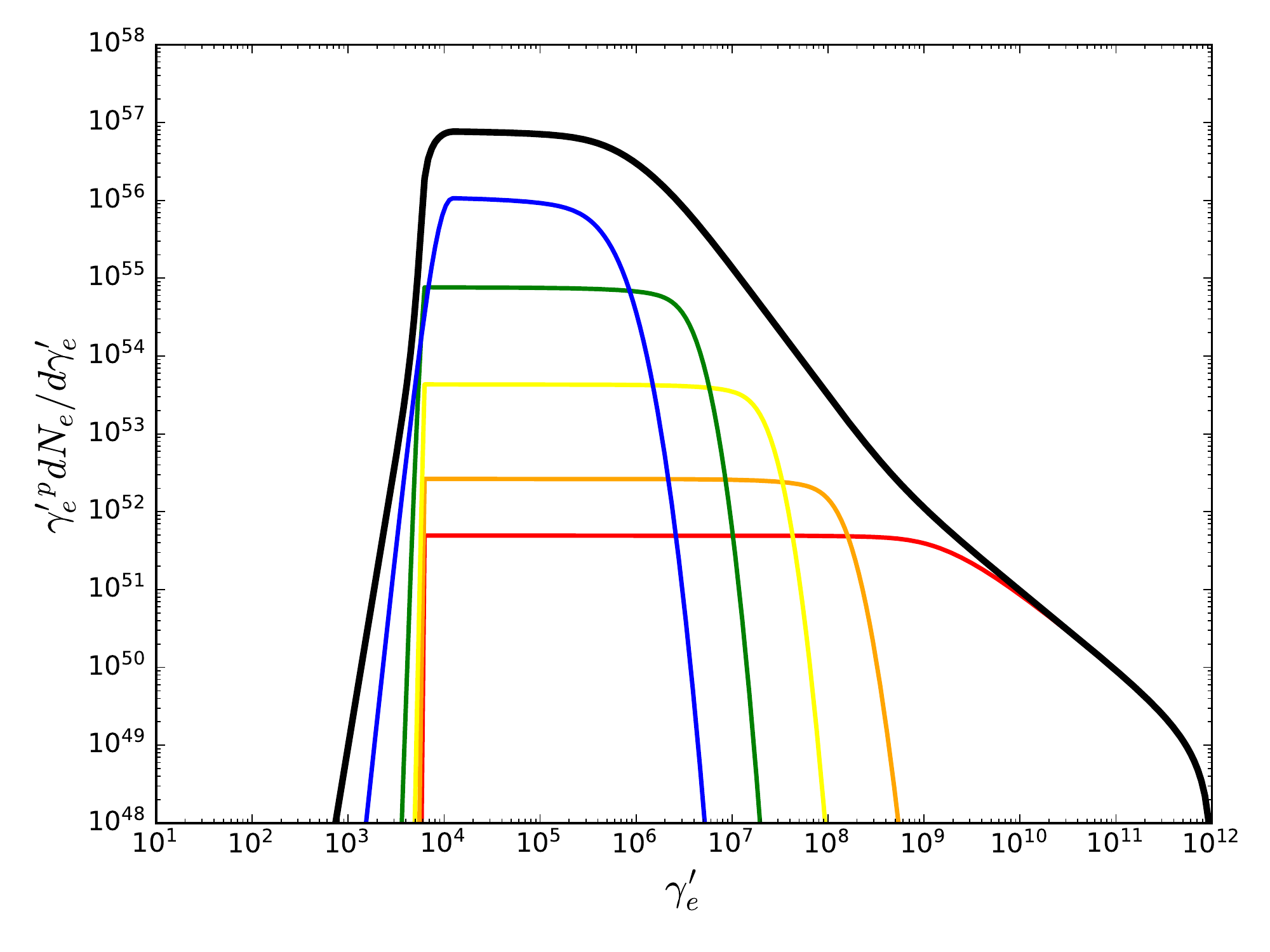}
\includegraphics[scale=0.4]{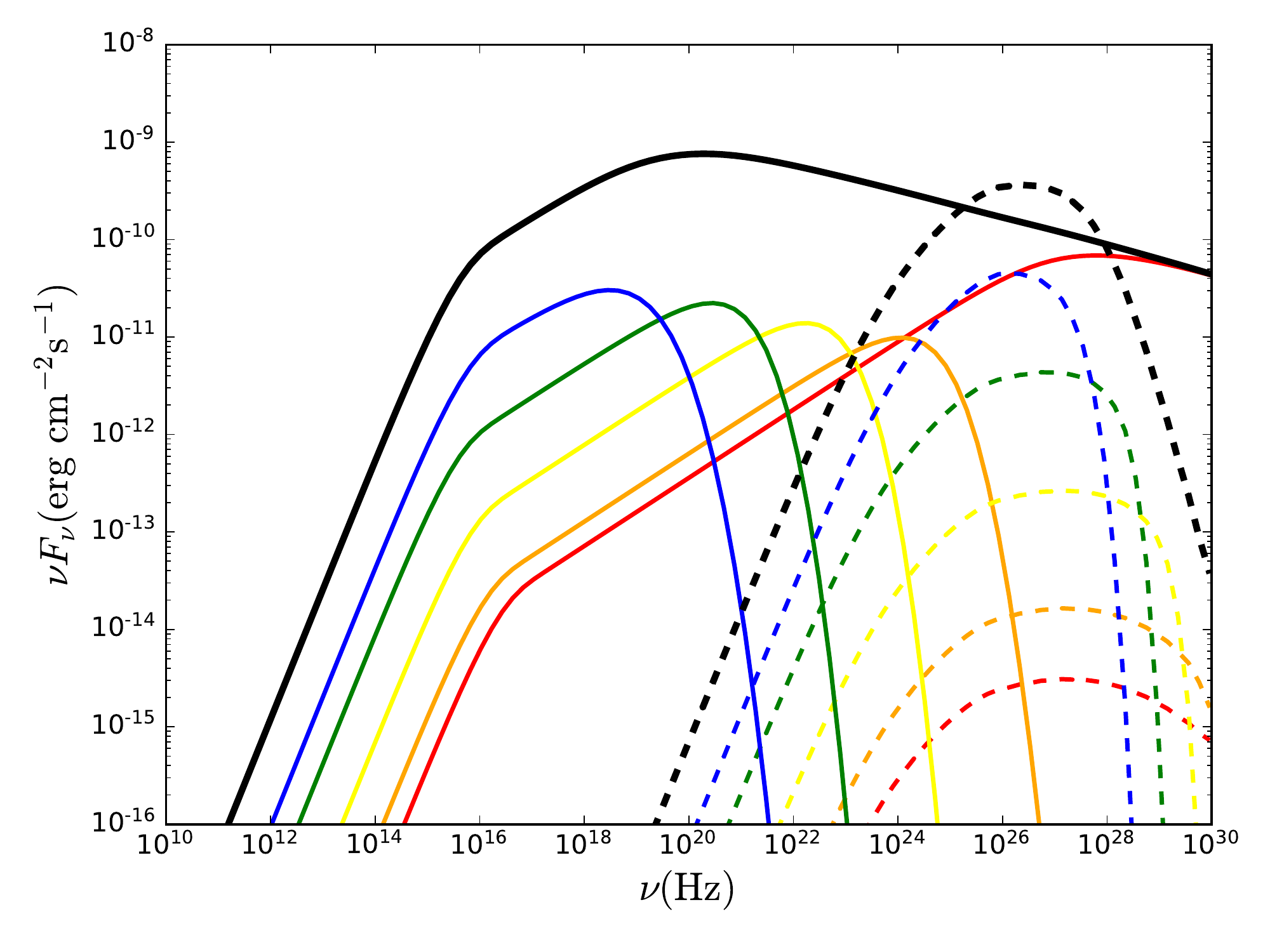}
\caption{An example to illustrate the decomposing of the injected electrons. Top left panel: the magnetic field strength versus proper time. The cases of some representative shells are marked with circles with different colors. The solid line corresponds to the magnetic field structure described by Eq. (\ref{B}). Top right panel: The synchrotron, IC and adiabatic cooling rates of the accelerated electrons in the representative shell as function of electron energy. Bottom left panel: The EDs of the representative shells and all downstream accelerated electrons (black line). Bottom right panel: The synchrotron (solid lines) and IC (dashed lines) spectra of the representative shells and all accelerated electrons (black line). The fiducial parameter values are adopted (see text).}
\label{Fig:numerical}
\end{figure*}

\begin{table*}
	\caption{The fiducial values of DM model parameters  used in the calculation.}
	\label{Tab:tab}
	\begin{tabular}{ccccccccccc}
		\hline
		 $\alpha_t$ & $\Delta_{\mu}$ & $\epsilon_{B+}$ & $\epsilon_{e}$ & $p$ & $\Gamma_0$ & $E$ $\rm [erg]$ & $n$ $\rm [cm^{-3}]$ & $t$ $\rm [s]$ & $z$ & $\gamma_{\max}'$\\
		\hline
		$-0.4$ & $10^2$ & $10^{-2}$ & 0.1 & 2.3 & 300 & $10^{53}$ & 0.1 & $10^2$ & 1 & $10^{12}$\\
		\hline
	\end{tabular}
\centering
\end{table*}

As mentioned above we decompose the accelerated electrons into a series of shells that injected into the shock right after the shock front in subsequent time intervals. We will calculate the temporal evolution of the ED and SED from each shell (see sections \ref{sec:ED} and \ref{sec:SED}). The total emission from the GRB afterglow at a given time will be the sum of the flux from all the individual shells at that time.

The magnetic field strength downstream but near the shock front is much stronger than that far away from the shock front due to the decay. The earlier injected electrons have flowed far downstream and stay in a region with decayed, small magnetic field, so they may contribute emission mainly in the low energy range; on the contrary, the lately injected electrons stay close to the shock front and make more contribution to high energy range. In order for accurately calculating the broadband afterglow SED, we should take care both the low and high energy range. For this reason, we adopt larger time intervals for the earlier injected shells but smaller intervals for later shells.  Within each time step, we calculate the total cooling rate via Eq.(\ref{gamma_dot}) and apply the fully implicit difference scheme (see \citet{chi99, Chang70}) to solve the continuity equation, Eq.(\ref{continuity equation}), for each shell, then we obtain the ED and the synchrotron and IC spectra in each time grid. In the following calculation of one SED, we adopt the total time grid number to be 100. To test the convergence of the numerical calculation, we try the other time grid number and find that if the total time grid number is $\gtrsim 100$, the numerical solution tends to be stable.

In the following investigation, a set of ficucial parameter values are considered, as shown in Tab.\ref{Tab:tab}. Fig.\ref{Fig:numerical} shows as an example the numerical treatment of the afterglow with the fiducial parameter values. 100 shells are injected into the downstream, then they cool and emit radiation. The top-left panel shows several representative shells' magnetic field as function of the shell's proper time. In our approach, the shells are well sampling the regions with decayed and undecayed magnetic fields. The other panels also show the cooling rates and the corresponding EDs and radiation spectra of the representative shells.


In the DM model, the KN suppression leads to an important dependence of the IC cooling rates on the electrons' Lorentz factor, which also modify the spectral shape of the synchrotron and IC components. In our code, we provide detailed calculations of IC spectrum accounting for the KN effect, as described in section \ref{lc cooling}. An approximation about seed photons for IC scatterings should be mentioned here. The quantity $n_{ \nu'}$ in Eq.(\ref{ssc}) is taken as the density of background synchrotron seed photons which are the sum of contributions from all shells injected in the earlier time steps. 
In fact, the electrons of a shell can only scatter those photons that arrive at the position of the shell. The synchrotron photons emitted from the earlier injected shells easily satisfy this requirement, since the convective velocity of the fluid is always smaller than $c$. However, the synchrotron photons emitted by the later injected electrons may not have time to reach the position of the shell being considered, except for those synchrotron photons emitted by later injected shells but not too far from the shell being considered. A precise calculation of $n_{\nu'}$ for given shell and given time should include contributions from only the shells from which the emitted photons can reach the given shell in time, i.e., both the relatively earlier injected shells and some of the relatively later injected shells in the vicinity of the certain shell. We compare the total ED and SED calculated by two algorithms of $n_{\nu'}$, one with the approximation and the other one with precise calculation, and find no significant difference between them. For simplicity, $n_{ \nu'}$ in Eq.(\ref{ssc}) is summing up contribution from all shells already injected into the shock downstream at the time concerned.

In order to test the validity of our code, we have compared our numerical results with those obtained by \citet{lemoine13b}, which derives the GRB afterglow spectra with DM in analytical and semi-analytical approaches. The results are shown in Appendix (\ref{test}). As one can see, the results of our numerical method are in general similar with the analytical results of \citet{lemoine13b}, verifying the correction of our algorithm and code. It should also be pointed out that the numerical results of DM model slightly exceeds the analytical ones at some energy ranges. This slight discrepancy may be due to the approximation in the analytical approaches. The numerical calculation follows the cooling and radiation of electrons and then their evolution in the DM model in details, whereas the analytical method somehow makes approximation that electrons given initial injected energy only stay and give rise to radiation in the downstream region where the proper convective timescale is comparable to the radiative cooling timescale or dynamical timescale.  

\section{Results}\label{results}

In this section, we will show the results of numerical calculation of the afterglow emission in the DM model especially, we compare the EDs and SEDs in the DM model with those in the HT model; discuss the effects of the magnetic field decay power-law exponent $\alpha_{t}$ and the undecaying characteristic scale $\Delta_{\mu}$ on EDs and SEDs; and calculate the temporal evolution of spectra and LCs in the DM model. We pay attention to the distinct characteristics of the DM model from the HT model. To avoid the effect of the electron energy cutoff on the spectrum, we will assume a large maximum energy of the accelerated electrons, $\gamma_{\rm max}'=10^{12}$, in this section. Without special mention, the fiducial parameter values are taken as Tab.\ref{Tab:tab}.
The SEDs will be presented in the observer's frame, and the EDs are presented in the rest frame of the postshock fluid.

\subsection{Spectra with DM in comparison with HT}\label{sec:DMvsHT}

The afterglow can be divided into two regimes, the fast and slow cooling regimes, corresponding to whether the bulk of the accelerated electrons can cool in a dynamical time of the shock. Define $\gamma_c'$ the cooling Lorentz factor with which the electron's cooling time is equal to the dynamical time. As the accelerated electron energy is dominated by low energy electrons (i.e., $p>2$), $\gamma_m'>\gamma_c'$ corresponds to the fast cooling regime, otherwise slow cooling regime. We compare the afterglow spectra between the DM and HT models in both fast and slow cooling regimes . 

\subsubsection{Slow cooling regime}

\begin{figure}
\vskip -0.0 true cm
\centering
\includegraphics[scale=0.4]{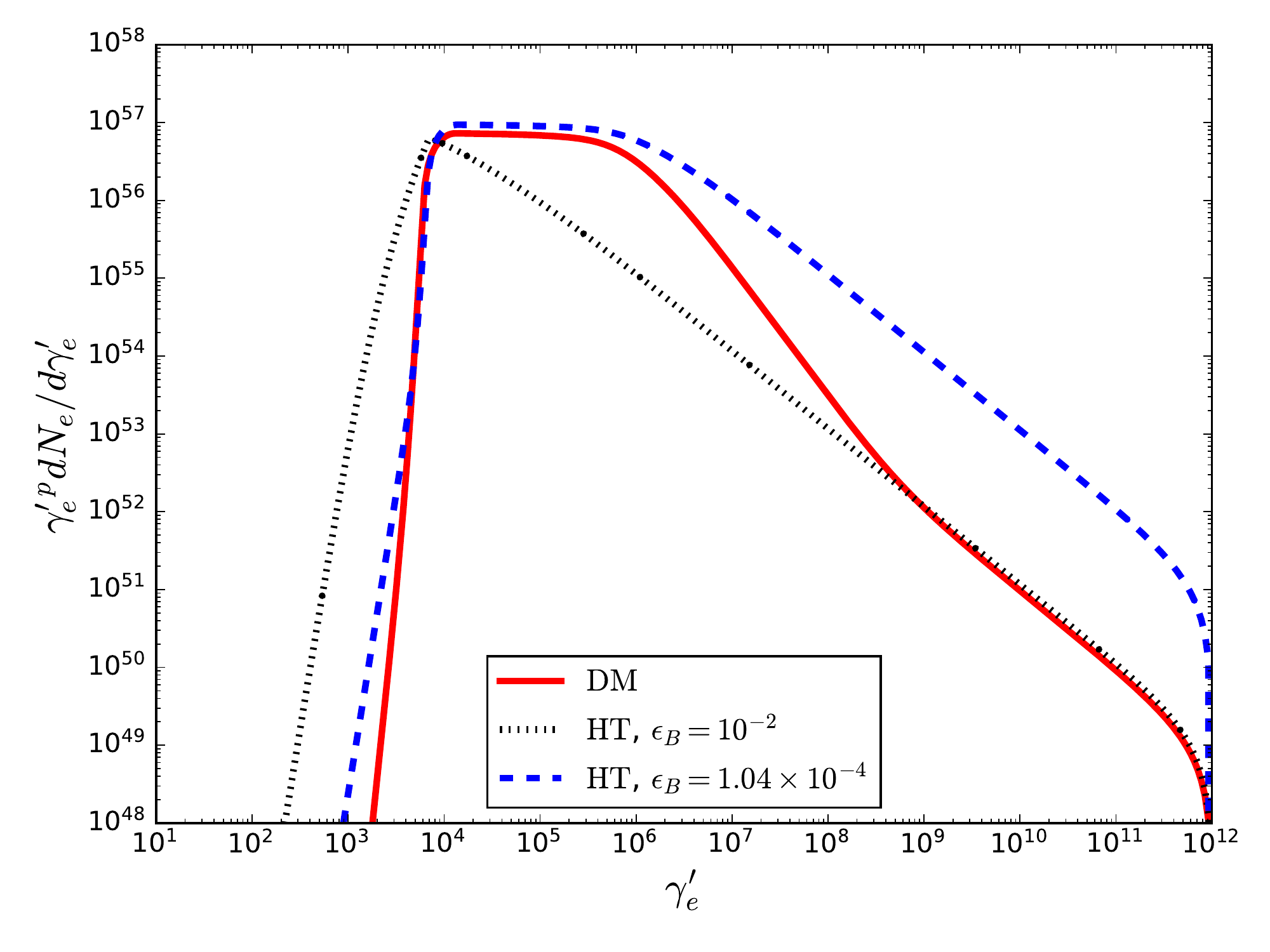}
\includegraphics[scale=0.4]{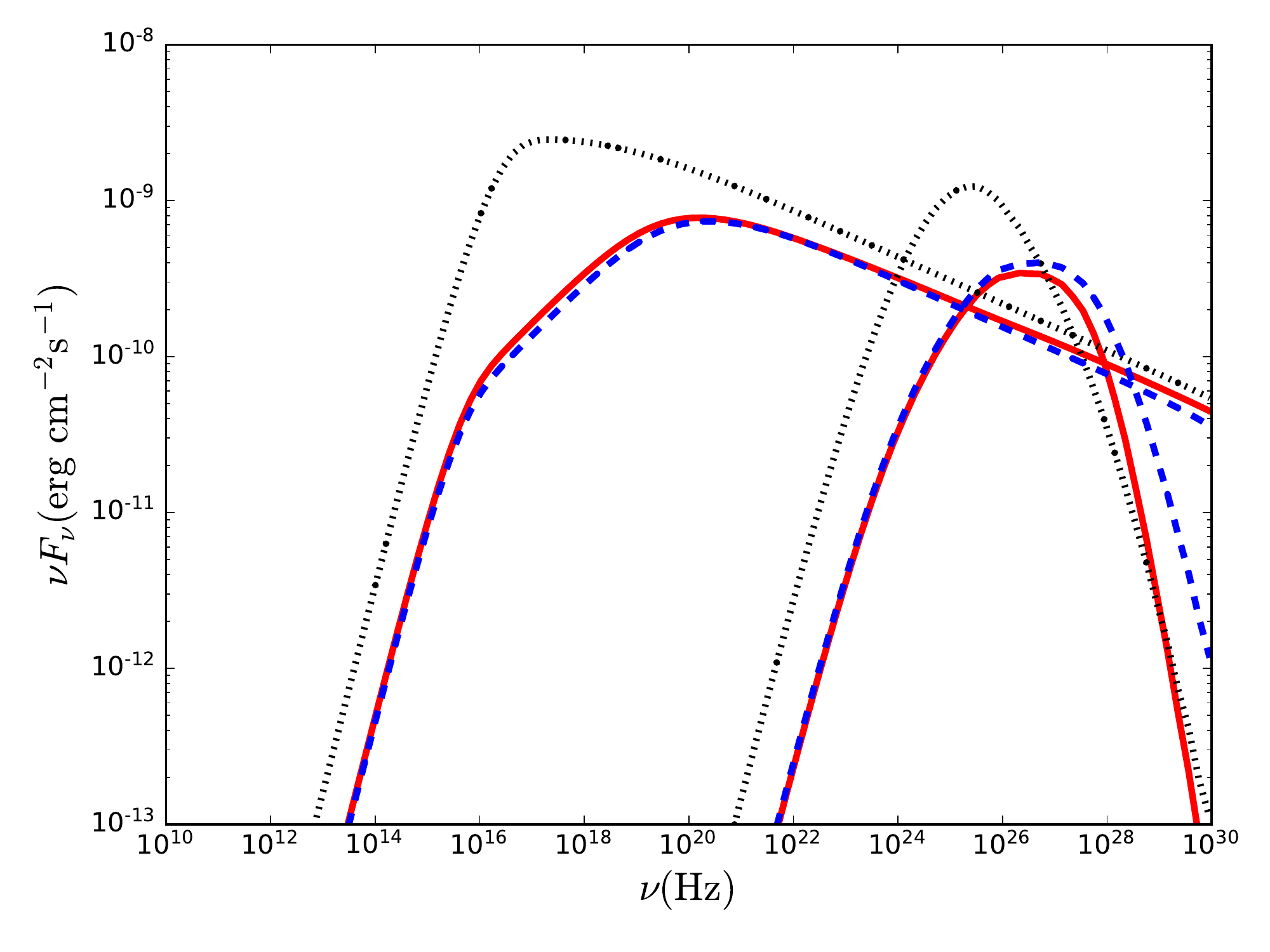}
\caption{
The ED (upper panel) and SED (lower panel; including synchrotron and IC spectral components) of the DM model in the slow cooling regime, in comparison with HT model. The DM model is presented by the red solid line, and the HT model is presented by the black dotted and blue dashed lines for two cases, $\epsilon_B=10^{-2}$ and $10^{-4}$, respectively. The other parameters are adopted with the fiducial values.}
\label{Fig:SED_slow}
\end{figure}

We show the results for the DM model taking the ficucial parameter values in Fig.\ref{Fig:SED_slow}, which is corresponding to the slow cooling regime. In order to compare DM and HT models, we take the same parameter values in them, except for the parameters describing the magnetic field. We consider two extreme values for the magnetic field equipartition parameter in HT model, which cover the range of the $\epsilon_B$ variation in the DM model: one is the largest allowed magnetic field strength in the DM model, i.e., the strength immediately behind the shock front, $\epsilon_B=\epsilon_{B+}$; the other is the possible lowest magnetic field strength in the DM model, i.e., the magnetic field in the fluid with the maximum proper time and the largest distance downstream away from the shock front, i.e., $\epsilon_B=\epsilon_B(t'_{\rm pM})$, where $t'_{\rm pM}$ is the maximum proper time of the shocked fluid.  In the calculation for the DM model with fiducial parameter values, we find  $\epsilon_B(t'_{\rm pM})= 10^{-4}$, which is then used in the HT model.

As slowly cooling electrons (i.e., the cooling time is larger than the dynamical time) keep the injected energy distribution, we show the ED in the upper panel by $dN_{e}/d\gamma_{e}'$ multiplied with $\gamma_{e}'^{p}$. Because electrons with higher energy usually cool faster, we generally expect high energy electrons mainly cool at the region closer to the shock front and with larger magnetic field. This is consistent with what is shown in the ED plot -- the ED of the low energy part in the DM model is consistent with the low magnetic field HT case, but the high energy end of the ED consistent with high magnetic field HT.

At low energies the flat ED segment clearly shows the slow cooling electrons, in both the DM and the HT with low magnetic field ($\epsilon_B=\epsilon_B(t'_{\rm pM})$). Note the HT with large magnetic field ($\epsilon_B=\epsilon_{B+}$) should be around the critical case that electrons start to cool significantly ($\gamma_{\rm min}'\simeq\gamma_c'$). We see that at $\gamma_e'\ga10^6$ the ED turns steeper, implying that the electrons become fast cooling. Indeed, from Fig. \ref{Fig:numerical} we see that the adiabatic cooling rate and the radiative cooling rate becomes equal around $\gamma_e'\sim10^6$ for the farthest fluid, implying $\gamma_c'\sim10^6$, because the adiabatic cooling time is comparable to the dynamical time.

At the high energy end, the ED becomes $dN_{e}/d\gamma_{e}' \propto \gamma_{e}'^{-p-1}$, which is expected for electrons fast cooling by synchrotron radiation. In fact Fig. \ref{Fig:numerical} shows that at high energies the IC cooling suffers strong KN suppression and is dominated by the synchrotron cooling, and that the synchrotron cooling dominates adiabatic cooling, implying that the electrons are fast cooling. The consistency with the HT with large magnetic field $\epsilon_B=\epsilon_{B+}$ suggests that the highest energy electrons mainly cool in the undecayed magnetic field region and the dominant cooling mechanism is synchrotron cooling. The ED break at $\gamma_e'\sim10^{9}$  corresponds to the synchrotron cooling time equal to the proper time in the undecayed magnetic field, $t'_{\rm \mu +}$. The part of ED between the $-p$ and $-p-1$ slope segments reflects electrons cool significantly at different regions with decayed magnetic field. Overall, the DM obviously change the ED, and make it deviate from the standard broken power-law.

The lower panel of Fig.\ref{Fig:SED_slow} shows the observed SEDs, including the synchrotron and IC components. The synchrotron components do not show a high energy cutoff in the plot because we adopt a very large $\gamma_{\rm max}'$. One sees that the synchrotron spectrum in DM model is very close to that in HT model of low magnetic field in a broad energy range, however the IC component is significantly different from the HT model even with low magnetic field. Thus, with DM the change of IC component is more obvious than the synchrotron one. With similar synchrotron spectra, the change in IC spectral profile can be easily understood by the change of the ED (the upper panel of Fig. \ref{Fig:SED_slow}).

It is seen that the low energy part of the IC spectral component is produced in the low magnetic field region $\epsilon_{B}=\epsilon_B(t'_{\rm pM})$, whereas the high energy part in the large magnetic field region $\epsilon_{B}=\epsilon_{B+}$. Basically, one may try to obtain two different $\epsilon_B$ values by using HT models to low energy part and high energy part of IC component, and the discrepancy between them may derive the magnetic field decay exponent $\alpha_t$. Moreover, one can see that the IC to synchrotron ratio in flux with DM is similar to the HT with low magnetic field, but larger than that of HT with large magnetic field. Thus, similar to the HT of low magnetic field, the afterglow with DM tends to produce strong IC spectral component.

\subsubsection{Fast cooling regime}

\begin{figure}
\vskip -0.0 true cm
\centering
\includegraphics[scale=0.4]{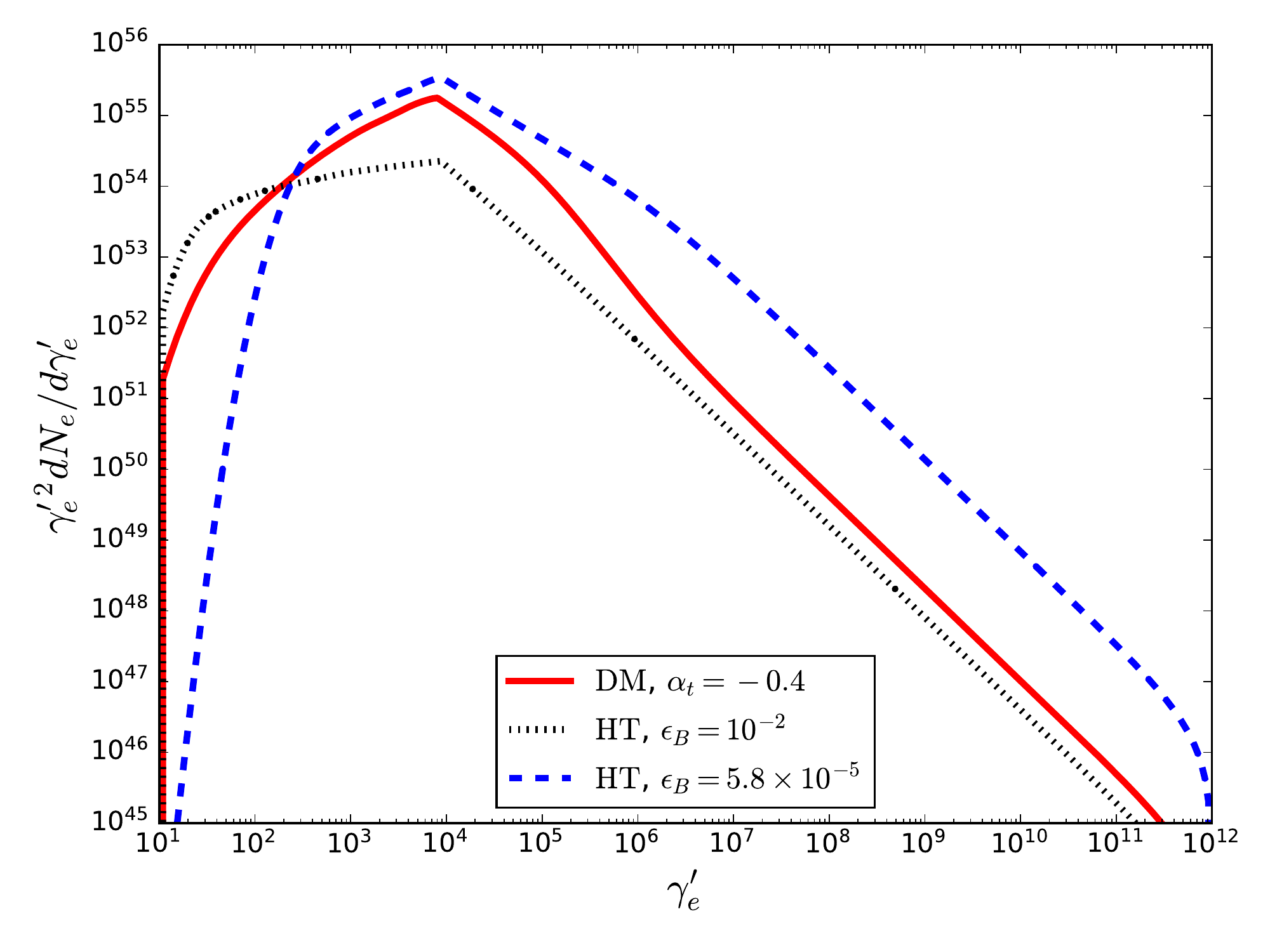}
\includegraphics[scale=0.4]{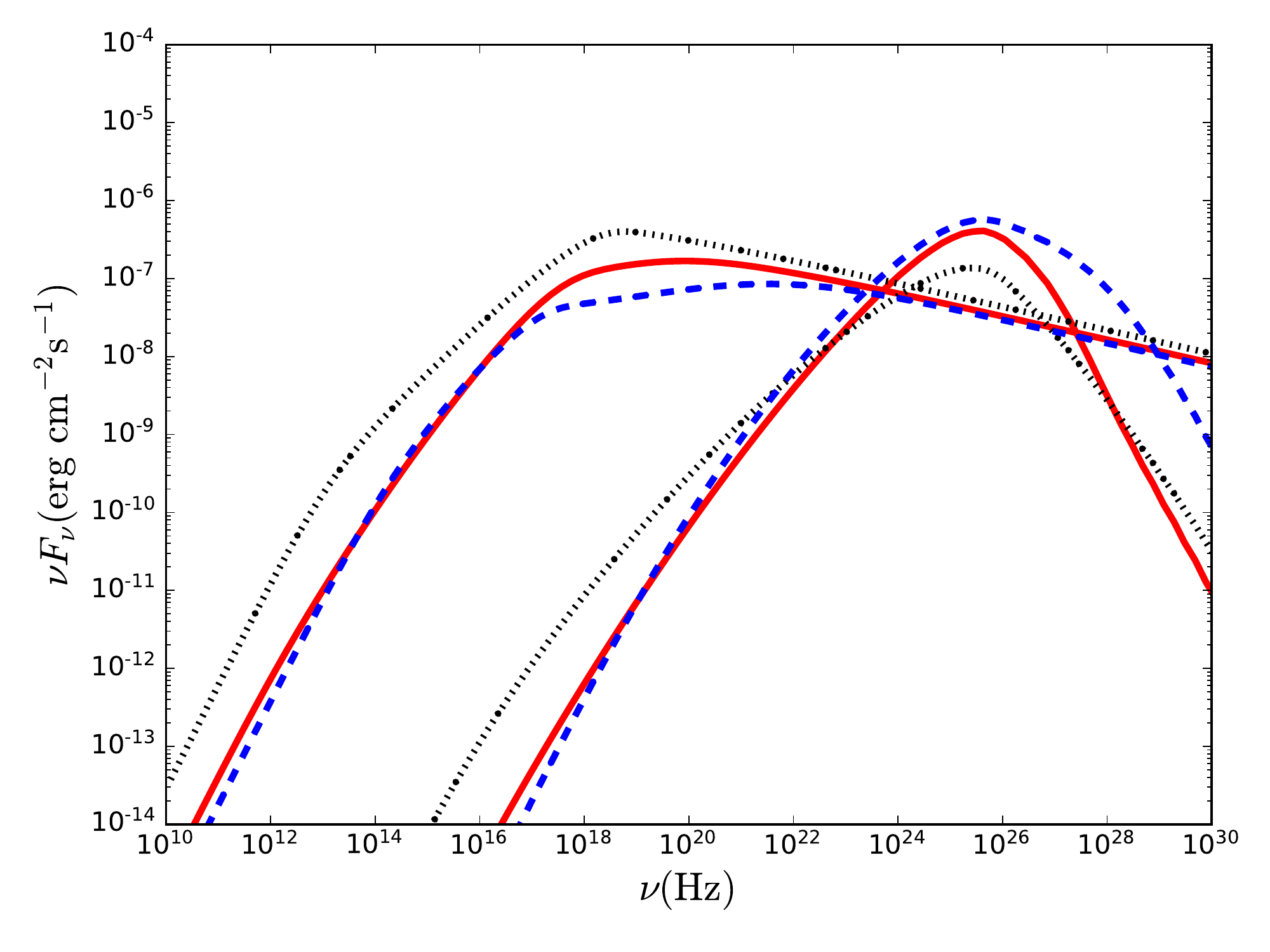}
\caption{Similar to Fig. \ref{Fig:SED_slow} but for the fast cooling regime. The HT models are for $\epsilon_B=10^{-2}$ (black dotted line) and $5.8 \times 10^{-5}$ (blue dashed line). The parameter values are $E=10^{54} \rm erg$, $n=10^2 \rm cm^{-3}$ and $t=10 \rm s$, and the others are adopted with the fiducial values.}
\label{Fig:SED_fast}
\end{figure}

We consider the DM model in the fast cooling regime by taking parameters $E=10^{54} \rm erg$, $n=10^2 \rm cm^{-3}$ and $t=10\rm s$, and the others with the fiducial values. In this case we obtain $\epsilon_B(t_{\rm pM})=5.8 \times 10^{-5}$. Fig.\ref{Fig:SED_fast} shows the DM in comparison with the two extreme HT cases. All three are in fast cooling regime. It is well known in the standard afterglow model, the case of extreme fast cooling regime will show an ED of $dN_{e}/d\gamma_{e}' \propto \gamma_{e}'^{-2}$, thus in the upper panel of Fig.\ref{Fig:SED_fast} the EDs shown have been multiplied by $\gamma_{e}'^{2}$.  The ED break corresponding to $\gamma_{\rm min}'$ appears around $\gamma_e'\sim 10^4$. 


For HT with low magnetic field, $\epsilon_B=5.8 \times 10^{-5}$, the high energy end is the standard fast cooling slope with $dN_{e}/d\gamma_{e}' \propto \gamma_{e}'^{-p-1}$, where the electron cooling is dominated by synchrotron cooling (with strong KN suppression on IC cooling). But below $\gamma_e'\sim 10^6$, the KN suppression starts to weaken and IC cooling becomes more important, so that the ED deviates from the $\gamma_{e}'^{-p-1}$ slope. For HT with large magnetic field, $\epsilon_B=10^{-2}$, only a $dN_{e}/d\gamma_{e}' \propto \gamma_{e}'^{-p-1}$ slope shows up in the high energy part because the IC cooling is not important and KN effect does not come in to play. Below the  $\gamma_{\rm min}'$ break is the low energy tail due to the fast cooling. Due to the KN correction of IC cooling, the EDs deviate from the slope of $-2$ and are different from each other between the two HT cases.

As for the DM case, from the upper panel of Fig.\ref{Fig:SED_fast}, we can see that the result of ED lies between the two extreme HT cases. Similar to slow cooling case, the very high energy part of the ED is consistent with that of HT model with large $\epsilon_B$, whereas the very low energy part becomes close to the HT model with low $\epsilon_B$. This is, again, because the highest energy electrons cool mainly in the large magnetic field region with $\epsilon_B=\epsilon_{B+}$, showing a $-p-1$ slope, whereas the lowest energy electrons cools mainly in the relatively low magnetic field region. The ED deviates from the $-p-1$ slope below $\gamma_e'\sim10^6$, which corresponds to the electrons with synchrotron cooling time in the undecayed magnetic field equal to $t'_{\rm \mu +}$.

The lower panel of Fig.\ref{Fig:SED_fast} provides the SEDs. The synchrotron component below the spectral peak is more consistent with the low $\epsilon_B$ HT case; on the other end all three cases are consistent with each other, because the high energy electrons lose most of their energy in synchrotron radiation, due to strong KN suppression of IC emission. As for the IC component, the part below the spectral peak is again more close to the low $\epsilon_B$ HT case. However, the IC emission above the spectral peak becomes closer to the large $\epsilon_B$ HT case, and the difference from the low $\epsilon_B$ HT case is obvious and the slope of IC spectrum becomes steeper than the HT model.


\subsection{Effect of magnetic field structure}
\subsubsection{Spectra with different $\alpha_t$}\

\begin{figure}
\vskip -0.0 true cm
\centering
\includegraphics[scale=0.4]{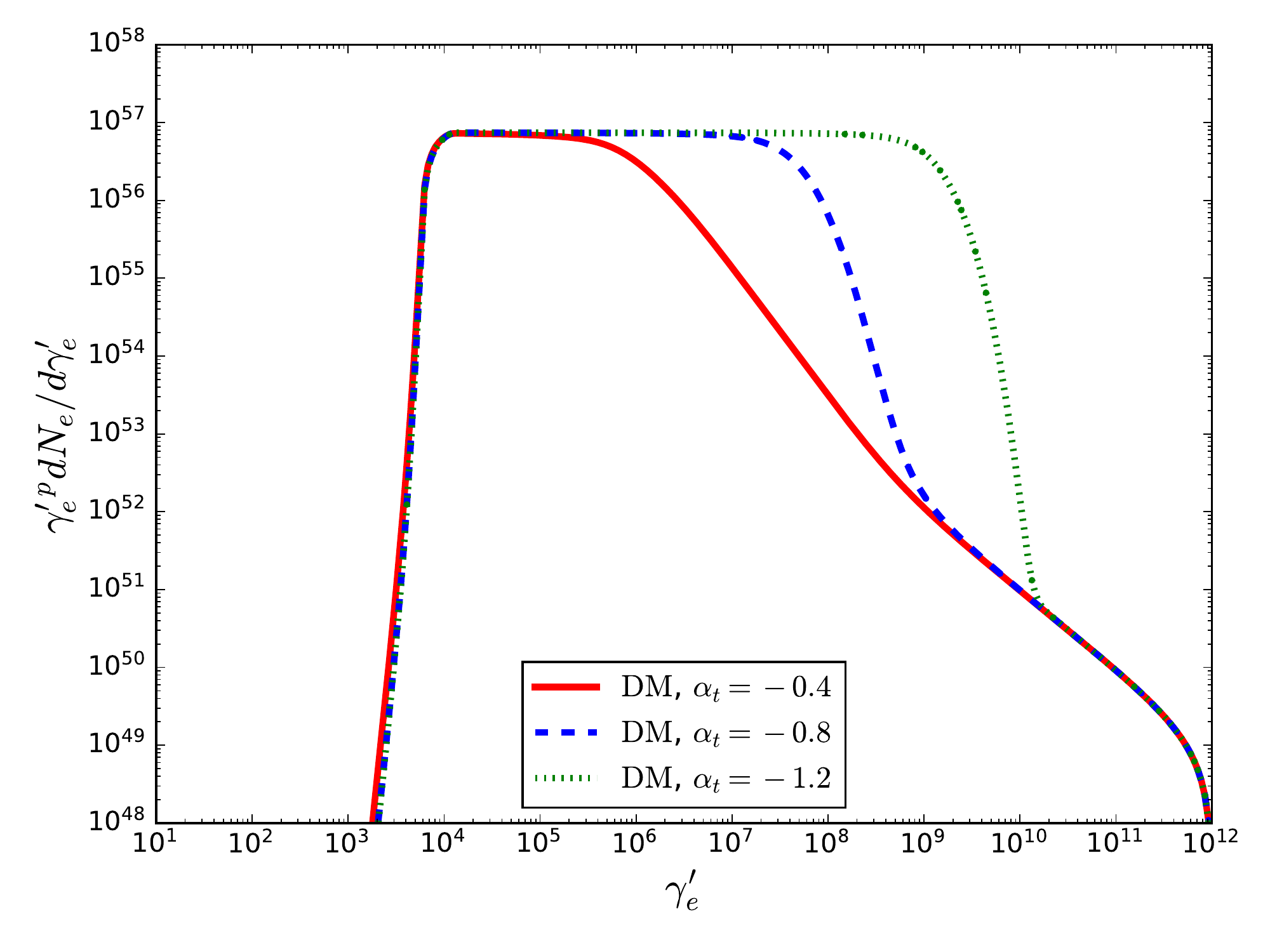}
\includegraphics[scale=0.4]{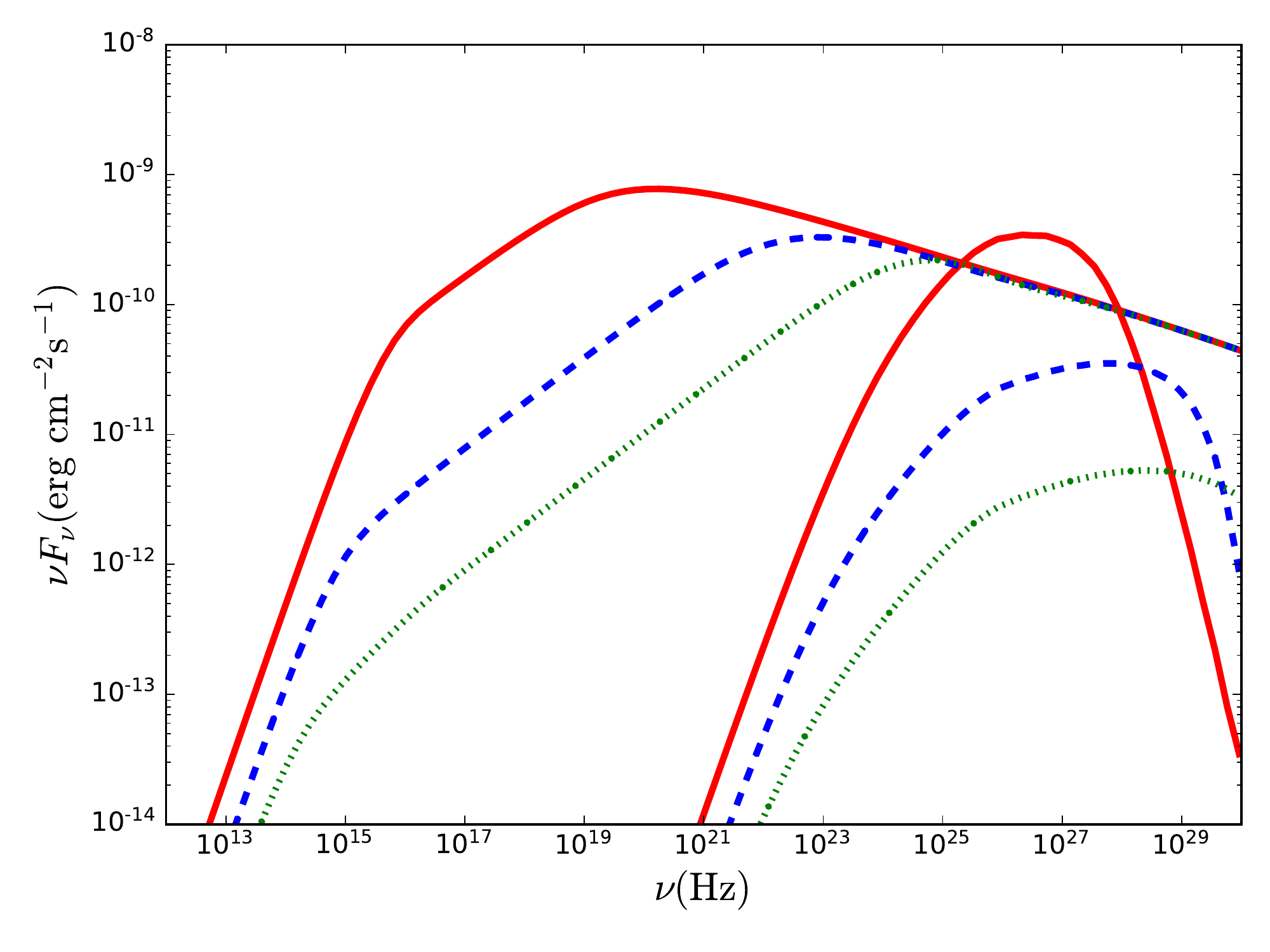}
\caption{EDs (upper panel) and SEDs (lower panel) of the DM model with different decay power-law exponents, $\alpha_{t}=-0.4$ (red solid lines), $-0.8$ (blue dashed lines) and $-1.2$ (green dotted lines). The other parameters are adopted with the fiducial values. }
\label{Fig:index}
\end{figure}

\begin{figure}
\vskip -0.0 true cm
\centering
\includegraphics[scale=0.4]{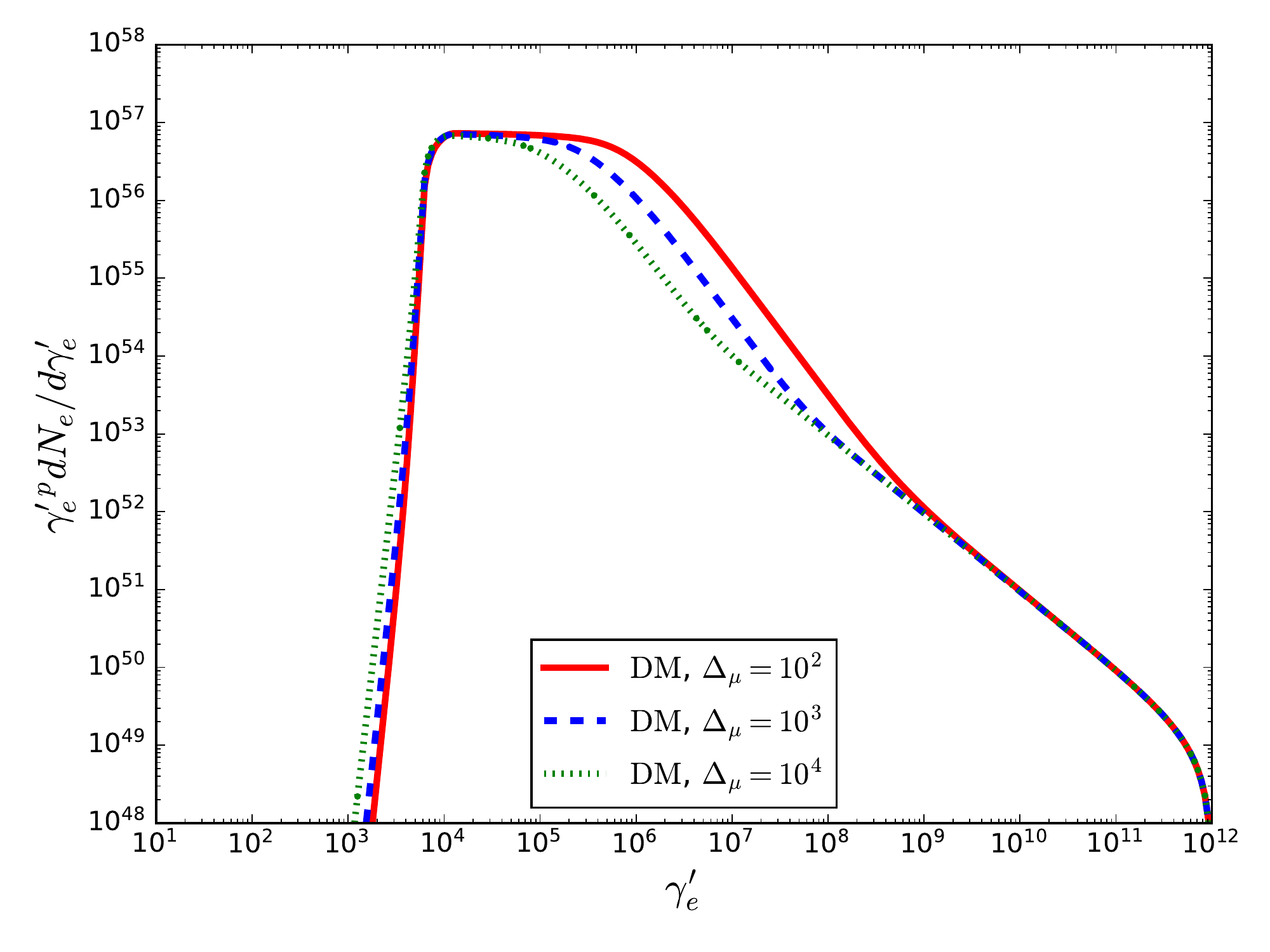}
\includegraphics[scale=0.4]{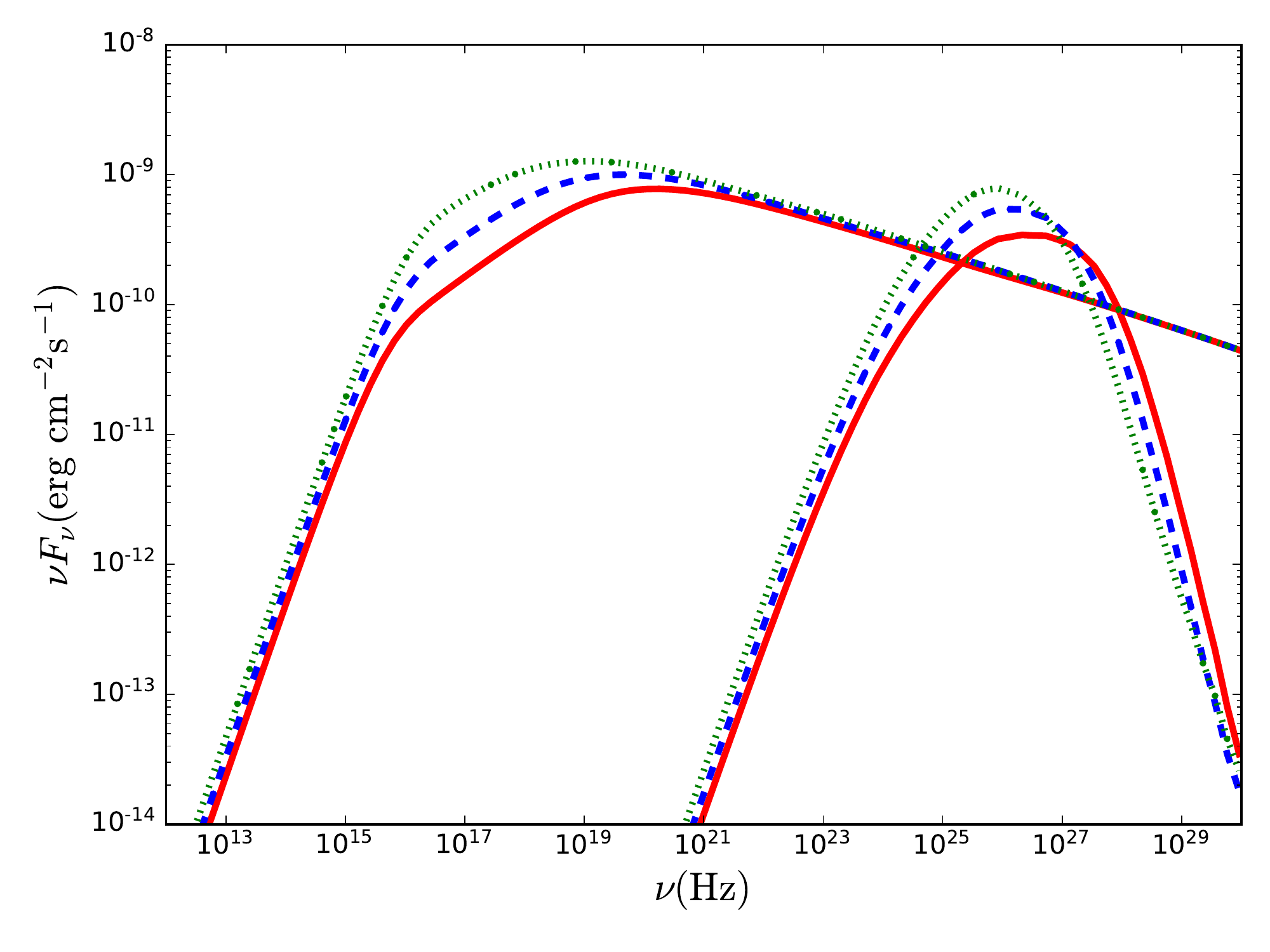}
\caption{Similar to Fig. \ref{Fig:index} but for different undecaying characteristic scale, $\Delta_{\mu}=10^2$ (red solid lines), $10^3$ (blue dashed lines) and $10^4$ (green dotted lines). The other parameters are adopted with the fiducial values. }
\label{Fig:delta}
\end{figure}

We investigate the changes of EDs and SEDs with different magnetic field decay power-law exponent $\alpha_{t}$, as shown in Fig. \ref{Fig:index}. With fiducial values, the afterglow is in the slow cooling regime.

From the upper panel of Fig.\ref{Fig:index}, first we can see the slow cooling segment in the EDs  corresponding to $\gamma_{e}'<\gamma_c'$, with the slope $-p$. The EDs in the cases of different $\alpha_t$ are coincident with each other. The ED deviates from slow cooling segment at $\gamma_c'$. For smaller $\alpha_t$, i.e., faster decay of the magnetic field, $\gamma_c'$ is larger, because faster decay results in lower magnetic field in the same proper time of the fluid (or the distance away from the shock front) and hence smaller cooling rate and larger cooling Lorentz factor of electrons. For smaller $\alpha_t$ the ED turns more sharply into the high energy end where the ED slope is $-p-1$, corresponding to the part of electrons fast cooling via synchrotron cooling (IC cooling is suppressed by strong KN effect at very high energies) in the high magnetic field region with $\epsilon_B=\epsilon_{B+}$. 

The lower panel of Fig. \ref{Fig:index} presents the results for the relevant SEDs. The bulk emission power is lower for faster magnetic field decay (smaller $\alpha_t$), since the magnetic field strength is lower and the synchrotron photon energy density is lower for IC emission. 

The synchrotron components in the SEDs show generally the standard spectral shape in slow cooling regime, with a low energy break relevant to electrons with $\gamma_{\rm min}'$ and the peak energy relevant to electrons with $\gamma_c'$. In some details, since the magnetic field decays to lower strength for smaller $\alpha_t$ cases, the low energy break and the synchrotron flux around the break is lower. At the highest energy end for all cases the electrons rapidly emit their energy in synchrotron photons, so the SEDs match with each other in different $\alpha_t$ cases. The SEDs in the middle correspond to the electrons fastly cooling in the region of the decay slope of the magnetic field, and for smaller $\alpha_t$ the turnover to the highest energy segment is sharper.  

 It appears also obviously that for smaller $\alpha_t$, the IC spectral component (the bump at higher energies) becomes relatively lower compared with the synchrotron spectral component (the bump at lower energies). The two cases with $\alpha_t=-0.8$ and $-1.2$ show IC component even below the synchrotron one (although assuming electron energy distribution extending to very high energy). This is in contrast with the general picture in standard afterglow models, where lower magnetic field cases give rise to relatively stronger IC emission. The reason here is due to the strong KN suppression on the IC emission. For lower $\alpha_t$ the synchrotron spectral peak around $\nu(\gamma_c')$ move toward higher frequency where the KN effect become more stringent, suppressing the IC emission. Thus to have stronger IC emission at high energies, faster decay of the magnetic field may not help due to the KN effect.

\subsubsection{Spectra with different $\Delta_{\mu}$}\

We investigate the changes of EDs and SEDs with different undecaying characteristic scale $\Delta_{\mu}$, as shown in Fig.\ref{Fig:delta}. With fiducial values, the afterglow is in the slow cooling regime. We consider three cases, i.e., $\Delta_{\mu}=10^2$, $10^3$, and $10^4$.

The upper panel of Fig.\ref{Fig:delta} shows the EDs with different $\Delta_{\mu}$. Similar with the case of different $\alpha_t$, for the slow cooling segment, the EDs in the cases of different $\Delta_{\mu}$ are coincident with each other at $\gamma_e' < \gamma_c'$, with the slope $-p$. For different $\Delta_{\mu}$ values, the slope of ED turns from $-p$ to a similar slope at $\gamma_e' \gtrsim \gamma_c'$, but $\gamma_c'$'s are different, i.e., smaller $\Delta_{\mu}$ gives larger $\gamma_c'$. This is because shorter undecaying characteristic scale results in lower magnetic field in the same proper time of the fluid and hence smaller cooling rate, and hence even higher energy electrons can cool fast. But $\gamma_c'$ is less dependent on $\Delta_{\mu}$, relative to $\alpha_t$. If $\Delta_{\mu}$ increases by an order of magnitude, $\gamma_c'$ only decreases by a half order of magnitude. Furthermore, the EDs with different $\Delta_{\mu}$'s turn to be coincident with each other at the high energy end showing an ED slope of $-p-1$.

The lower panel of Fig.\ref{Fig:delta} shows the relevant SEDs.
The bulk emission power is lower for smaller $\Delta_{\mu}$, since the magnetic field strength is lower. The shapes of synchrotron and IC spctra with different $\Delta_{\mu}$ are similar, and show generally the standard spectral shape in slow cooling regime. For smaller $\Delta_{\mu}$, the low energy break related to $\gamma_m'$ is lower, while the peak energy related to $\gamma_c'$ is larger. At the high energy end, the SEDs of different $\Delta_{\mu}$ merge. General speaking, for different $\Delta_{\mu}$ values, the difference in SEDs is small, only within half an order of magnitude between the three cases. 

In short, different $\Delta_{\mu}$ values result in different magnetic field strength in the decayed part for the same proper time of the fluid, then the change of $\Delta_{\mu}$ affects the GRB afterglow emission, but the effect is less important than that of $\alpha_t$.

\subsection{Temporal evolution}\label{lc}
\subsubsection{Temporal SED evolution}\

\begin{table}
	\centering
	\caption{Parameter values Used in the HT models in Fig. \ref{Fig:tobs} for comparison with the DM model.}
	\label{Tab:tab1}
	\begin{tabular}{cc}
		\hline
		 $t$ [s] & $\epsilon_{B}$\\
		\hline
		$10$ &  $2.3 \times 10^{-4}$ \\
         $10^2$ & $1 \times 10^{-4}$ \\
         $10^3$ & $5.8 \times 10^{-5}$ \\
         $10^4$ & $3.3 \times 10^{-5}$\\
		\hline
	\end{tabular}
\end{table}

\begin{figure}
\vskip -0.0 true cm
\centering
\includegraphics[scale=0.4]{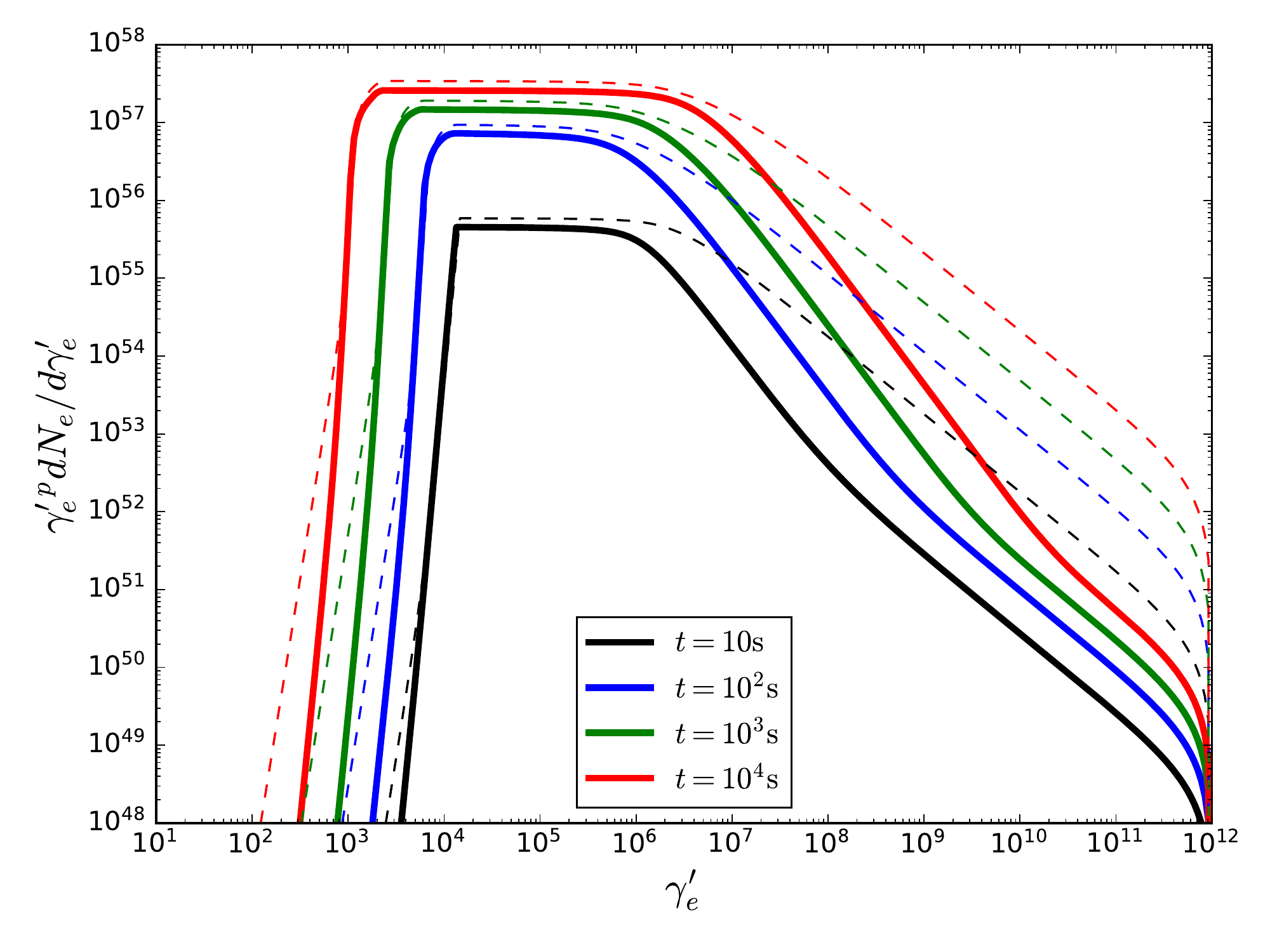}
\includegraphics[scale=0.4]{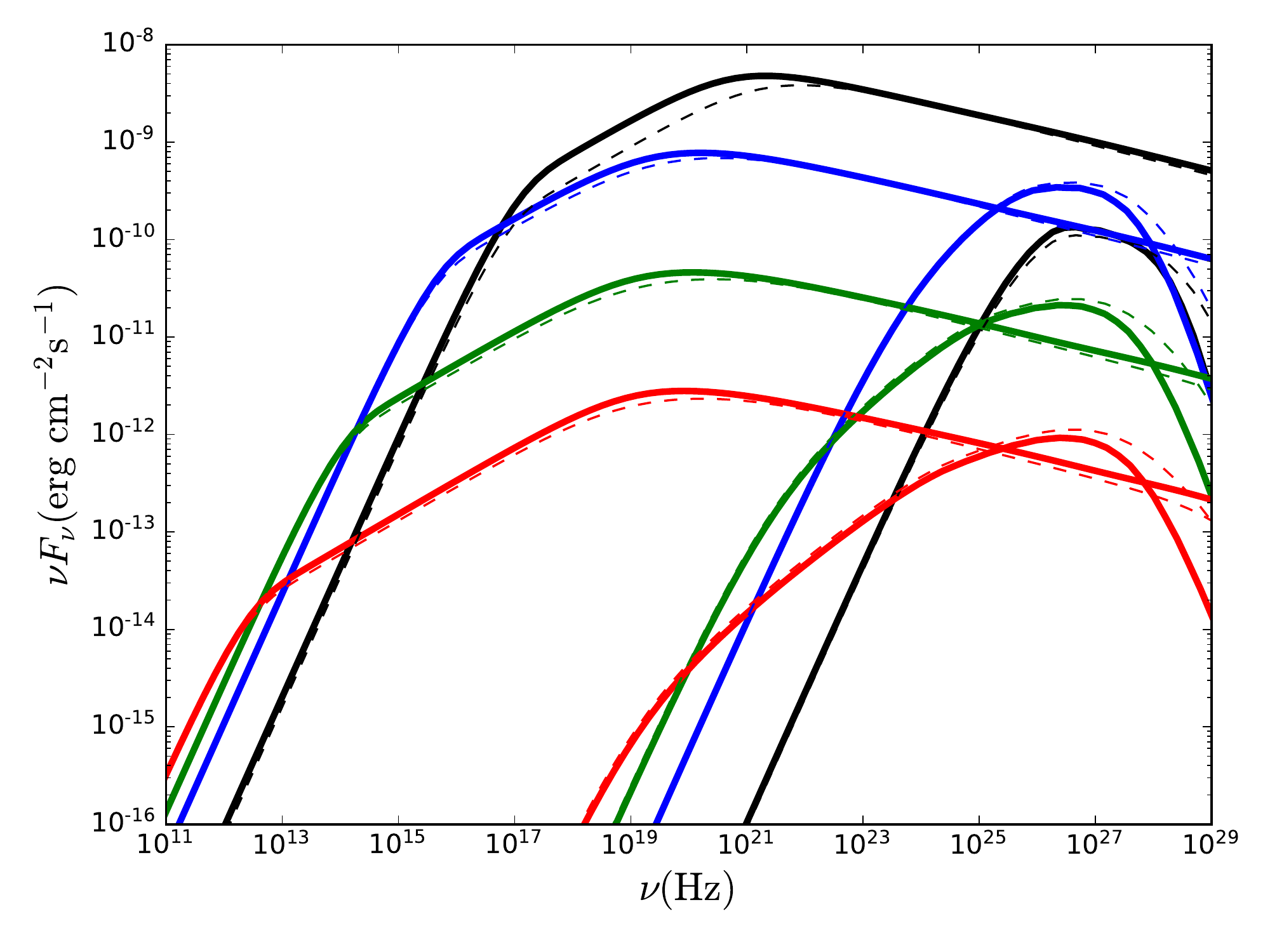}
\caption{The temporal evolution of EDs (upper panel) and SEDs (lower panel) in the DM model. The observer times are $t=10 \rm s$, $10^2 \rm s$, $10^3 \rm s$ and $10^4 \rm s$ for red, blue, green and black lines, respectively. The thick solid lines present the DM model, and for comparison, the thin dashed lines present different HT models with different observer times and $\epsilon_B$ (see Tab. \ref{Tab:tab1}). The other parameters are adopted with the fiducial values.}
\label{Fig:tobs}
\end{figure}

\begin{figure*}
\vskip -0.0 true cm
\centering
\includegraphics[scale=0.4]{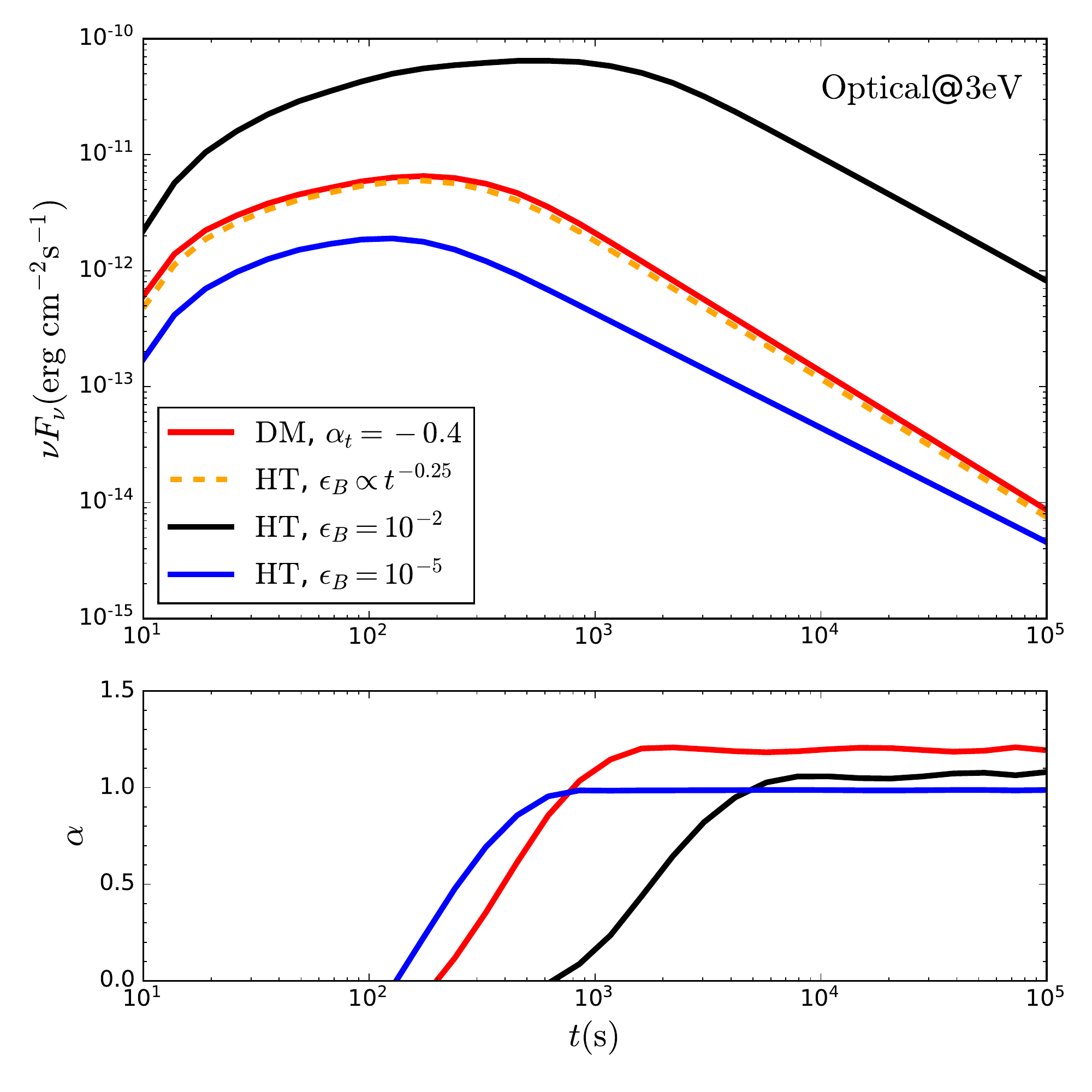}
\includegraphics[scale=0.4]{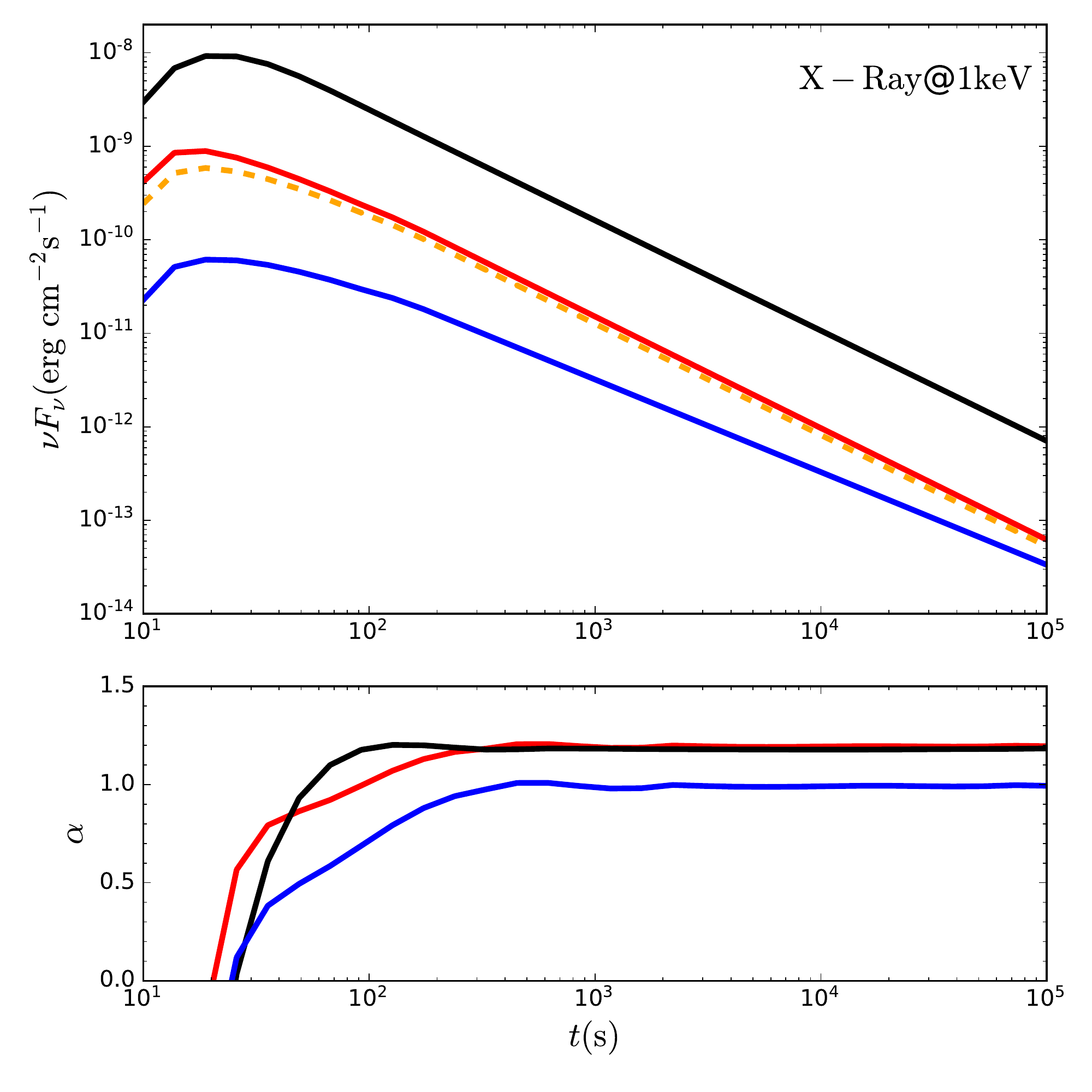}
\includegraphics[scale=0.4]{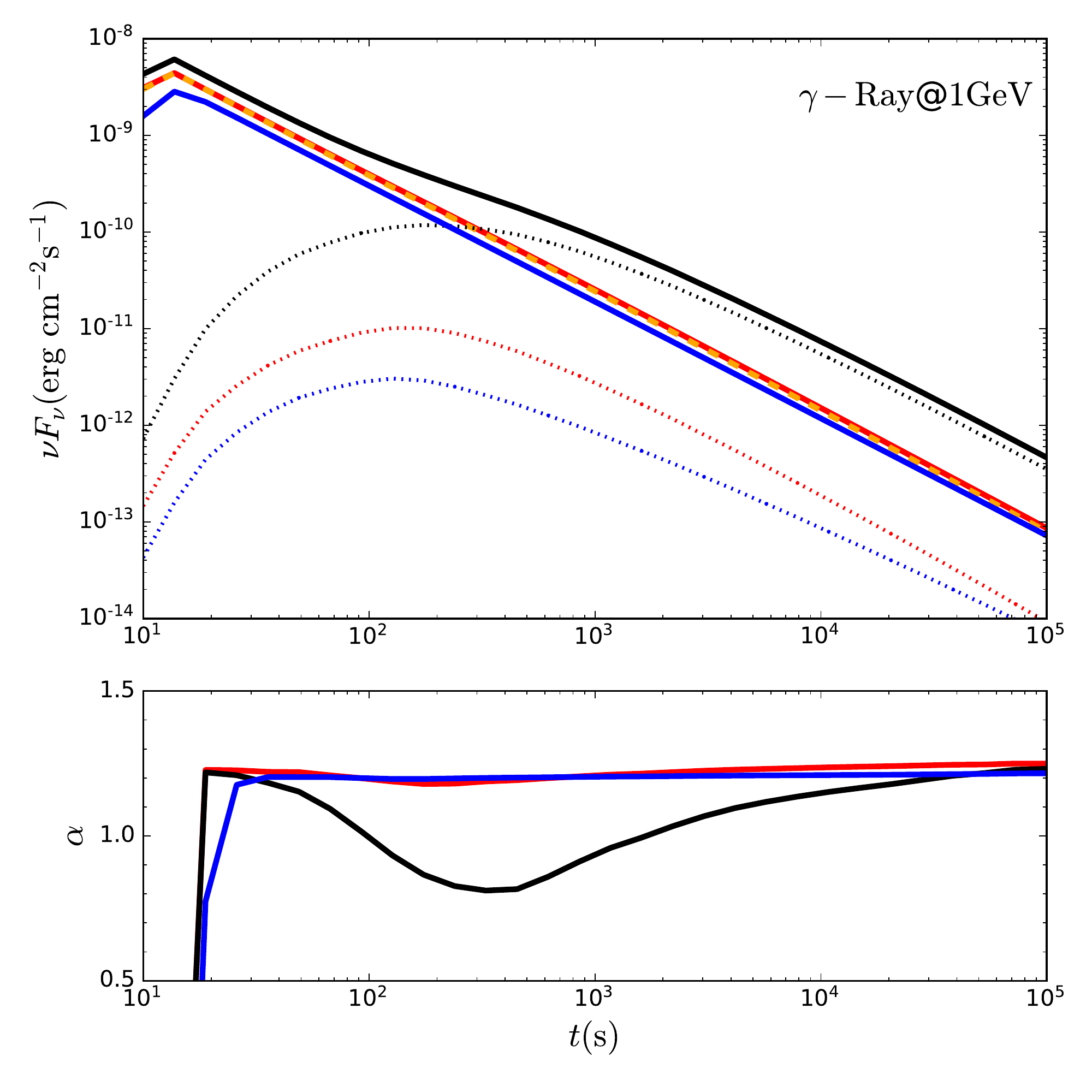}
\includegraphics[scale=0.4]{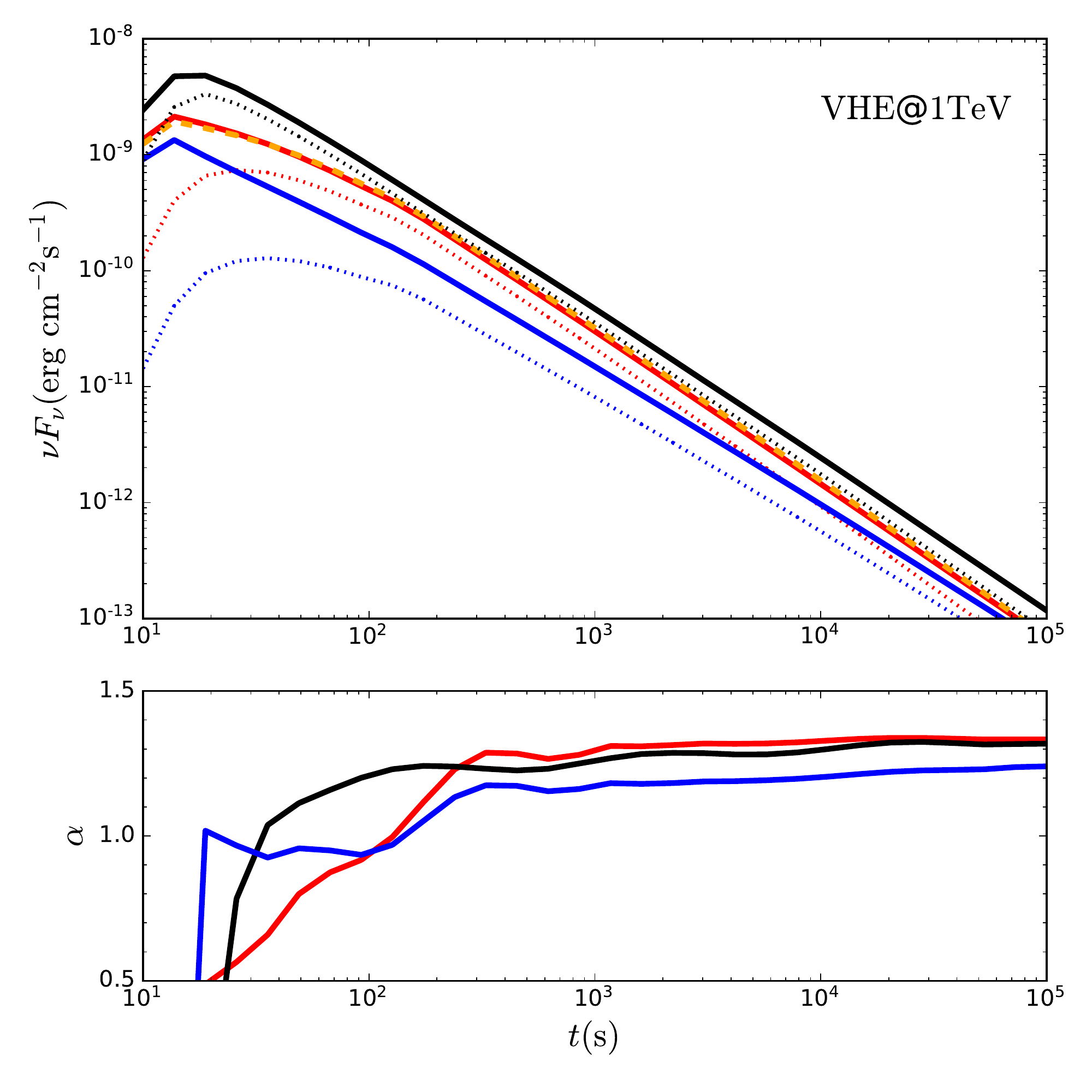}
\caption{Multi-band LCs (upper panel) and the temporal index evolution (lower panel) for DM and HT models. The red solid lines present the DM model with $\alpha_t=-0.4$, the orange dashed lines present the $\epsilon_B$-evolving HT model with $\epsilon_B\propto t^{-0.25}$, and the black and blue solid lines present the HT models with $\epsilon_B=10^{-2}$ and $10^{-5}$, respectively. The dotted lines are contribution from the IC emission. The other parameters are adopted with the fiducial values.}
\label{Fig:LC}
\end{figure*}

We investigate the temporal evolution of the afterglow spectra in DM model. Fig.\ref{Fig:tobs} shows the EDs and SEDs of DM model for different observer times. The first epoch with $t=10$s (presented in black lines in the figure) is during the coasting phase when the shock does not decelerate yet. Due to continuous injection, the electron number is increasing with time. The low energy cutoff $\gamma_{\rm min}'$ decreases with time due to the shock deceleration. 

For comparison we also show the HT models for the same observer times, but the HT models for different observer times are with different $\epsilon_B$ values, which are taken to be a low value, $\epsilon_B=\epsilon_B(t_{\rm pM}'[t])$ in the DM model, i.e., the equipartition value for the farthest downstream electrons at the observer time $t$. The $\epsilon_B$ values are given in Tab.\ref{Tab:tab1}. As discussed in section \ref{sec:DMvsHT} the high energy ED differs from the low $\epsilon_B$ HT models since the highest energy electrons cool rapidly in the large magnetic field region right behind the shock front.

The lower panel of Fig.\ref{Fig:tobs} shows the corresponding SEDs. We can see that the synchrotron spectra of DM model are in general consistent with the HT model for the same time and adopting the corresponding low $\epsilon_B$ values. But the IC component above the IC emission peak is somewhat significantly different from the relevant HT models, reflecting the ED difference at highest energies. 

The $\epsilon_B$ values as function of observer time in these HT models can be fit with a power law $\epsilon_B(t_{\rm pM}'[t])\propto t^{-l}$. One obtains $l=0.25$. This scaling can be explained below. For the earliest injected electrons, thus being farthest away from the shock front at certain time, the proper time since the injection is equal to the downstream comoving time $t_{\rm pM}'\simeq t'$. By Eq. (\ref{dynamic}) and Eq.(\ref{t_co}), one can derive that $t' \propto t^{5/8}$. Thus

\begin{equation}
\epsilon_B(t_{\rm pM}') \propto t_{\rm pM}'^{\alpha_t} \propto t^{5 \alpha_t /8}.
\end{equation}
Remind that $\alpha_t=-0.4$, we have $\epsilon_B(t'_{\rm pM}) \propto t^{-0.25}$.


\subsubsection{Multi-band LCs}\

We further investigate the multi-band LCs of the DM model.
In Fig.\ref{Fig:LC}, we compare the LCs and the temporal index $\alpha$ ($F_{\rm \nu} \propto t^{-\alpha}$) between the DM and HT models. We first compare DM model (the red lines) with the HT model with constant $\epsilon_B$. The black and blue lines correspond to the cases of $\epsilon_B=10^{-2}$ and $10^{-5}$, respectively. In all bands the LCs of DM model lie in the region between the two HT models. The decay part of the LCs in DM model usually shows slightly steeper slope than HT models, with a slightly larger temporal index $\alpha$.  

Next we consider a special HT model with $\epsilon_B$ evolving with time $t$, as motivated by the general consistency between the DM model spectra and that of the HT model but with different $\epsilon_B$'s (Fig \ref{Fig:tobs}). We consider a HT model where the postshock magnetic field equipartition parameter varies with observer time $t$, following the same value and evolution of the most early injected fluid in the DM model, i.e., the fluid relies at the farthest end of the downstream. For the case with $\epsilon_B \propto t^{-0.25}$, the calculated LCs are shown in Fig.\ref{Fig:LC} with orange dashed lines. We can see that the DM model and the $\epsilon_B$-varying HT model are matching each other very well. So we can conclude that the DM model emission can be well approximated by the HT model emission with $\epsilon_B\propto t^{5\alpha_t/8}$.

This can partly explain why the LCs in the DM model is somewhat steeper than the HT model with constant $\epsilon_B$. As $\epsilon_B$ decaying to lower values, the LC tends to turn from the high $\epsilon_B$ case with large flux to the low $\epsilon_B$ case with low flux. Thus the LC slope should be somewhat steeper than the HT models.

\section{Application to GRB 190114C}\label{apply}

\subsection{Observations}\

GRB 190114C was detected by Fermi-GBM with a duration of $T_{90}=116 \rm s$ \citep{fra2019a, rav2019}, and an isotropic equivalent energy and a luminosity of $E_{\rm iso} \sim \rm 3 \times 10^{53} erg$ and $L_{\rm iso} \sim 10^{53} \rm erg~s^{-1}$, respectively. Fermi-LAT detected high energy gamma-rays up to $\rm 22.9 ~ GeV$, whereas the Major Atmospheric Gamma Imaging Cherenkov (MAGIC) observed gamma-rays above 300 GeV with significance $>20\sigma$ at 50s after the trigger, making it the first ever GRB detected at TeV energies \citep{magic2019a, magic2019b, zhang2019}.

After the prompt burst, the gamma-ray LC shows a smooth decay, $\propto t^{-1.1}$ \citep{ajello2020, rav2019}. The X-ray and sub-TeV LCs also show power law decays, with the temporal indices  $\alpha_X \approx -1.36 \pm 0.02$ and $\alpha_{\rm TeV} \approx -1.51 \pm 0.04$, from XRT and MAGIC observations, respectively \citep[][the TeV LC is already obtained after correcting for attenuation by the extragalactic background light (EBL)]{magic2019a, magic2019b}. These smooth decays is well consistent with decelerating afterglow shock model. The NIR-optical behaviour is somewhat more complex \citep{magic2019a, fra2019b, laskar2019}, but except for an early fast decay, which can be accounted for by reverse shock emission, the LC also shows a shallower decay at $10^3 -10^5 \rm s$, similar to the other bands, until $t \gtrsim 10^5 \rm s$ where a steeper decay occurs, maybe due to the jet break effect.



\subsection{Modeling results}\

The radio to TeV emission from GRB 190114C has been explained to be the synchrotron and IC radiation from the GRB afterglow shock \citep{magic2019a, magic2019b, zhang2019, zhanghao2019, misra2019, fra2019b, derishev2019, wangxy2019}. The significant TeV emission has been mainly explained to be the IC emission from the afterglow. We here also apply the DM model to GRB 190114C. 

The afterglow shock deceleration time is corresponding to the observed peak time of the LC, $t_{\rm dec}(1+z)\simeq t_{\rm pk}$. With $z=0.4245$, $t_{\rm pk} \sim 10 \rm ~s$ \citep{magic2019a, magic2019b, ajello2020, rav2019} and using Eq.(\ref{t_dec}), we can obtain the initial bulk Lorentz factor $\Gamma_0 \sim$ a few 100's for  $E\sim10^{54} \rm~erg$ and $n\sim0.1-0.01 \rm~cm^{-3}$.

\begin{figure}
\vskip -0.0 true cm
\centering
\includegraphics[scale=0.4]{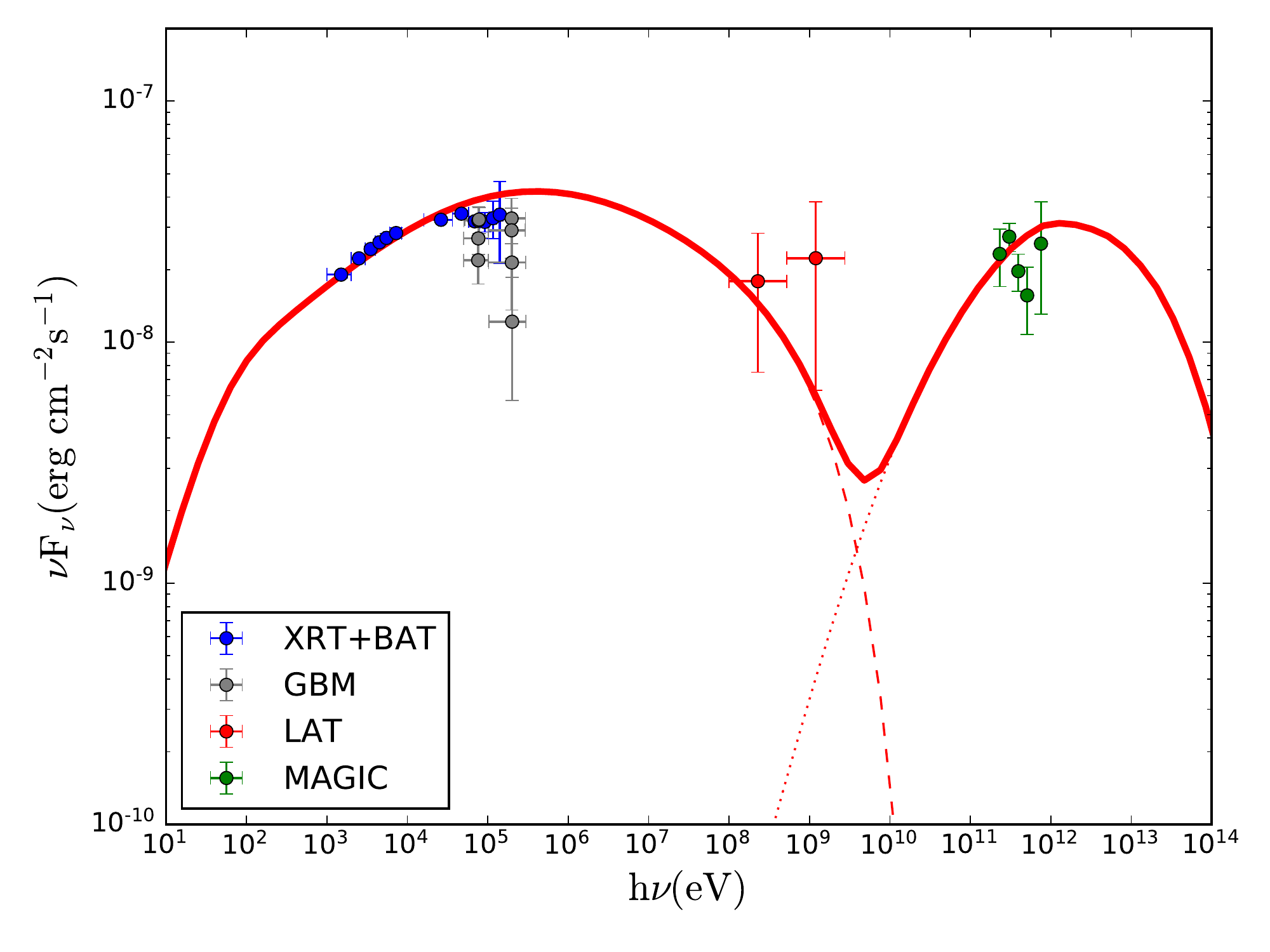}
\includegraphics[scale=0.4]{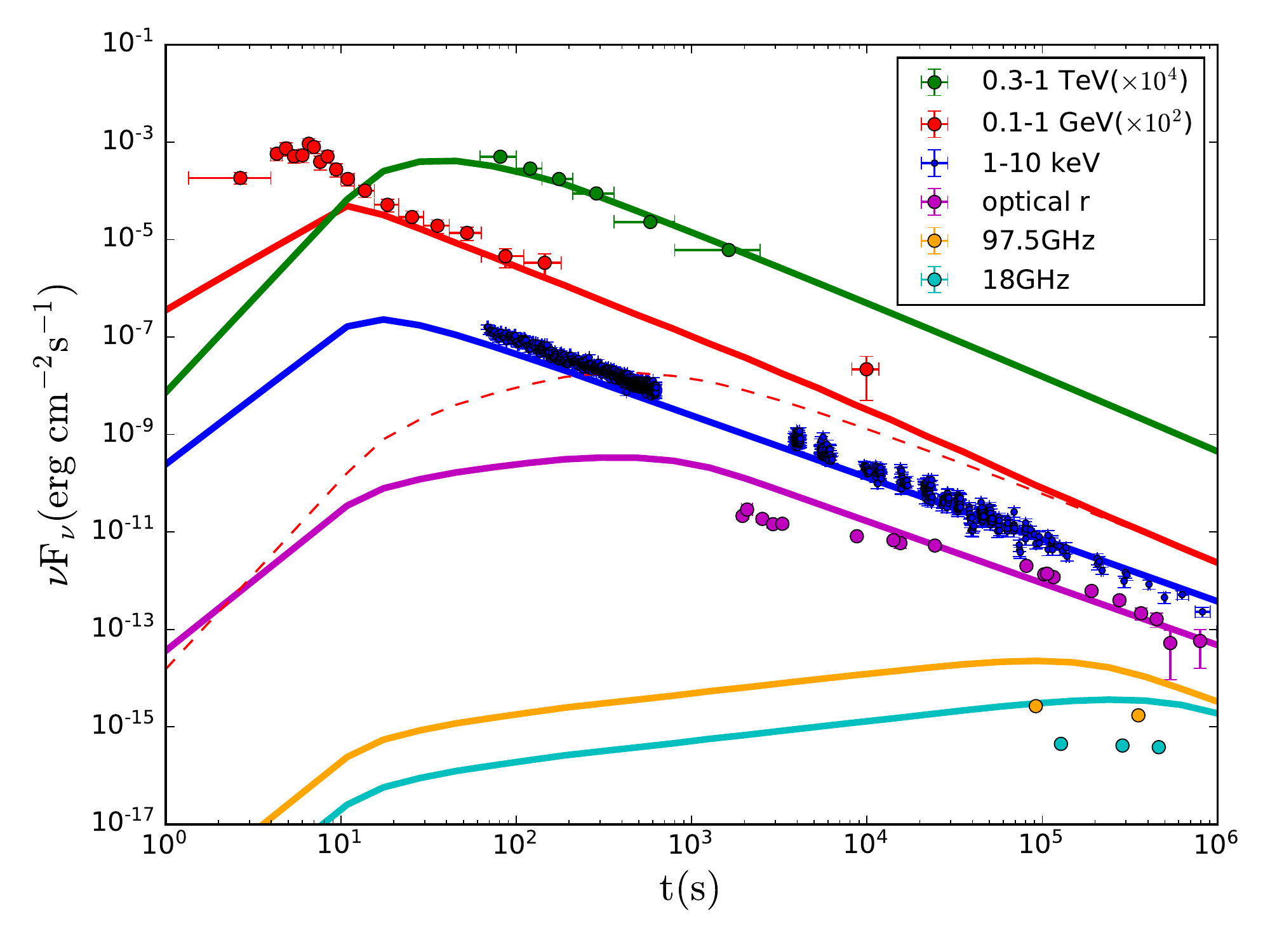}
\caption{The modeling of the multi-band data of GRB 190114C with the DM afterglow model. Upper panel: the broadband spectra in the time interval of 110-180s after the GRB trigger. The red dashed and dotted lines are the synchrotron and IC components, respectively, and the solid line is the total spectrum.  Lower panel: the multi-wavelength LCs from radio to TeV energy ranges. The red dashed line shows the IC component in the 0.1-1 GeV range. All the spectral and light curve data are taken from \citet{magic2019a} and references therein. The 0.3-1 TeV data have been corrected for EBL attenuation. The adopted parameter values in the DM model are shown in Tab.\ref{Tab:tab_190114c}.}
\label{24fitting190114}
\end{figure}

In modeling the data, the maximum Lorentz factor of accelerated electrons, $\gamma_{\rm max}$, should be determined by Eq.(\ref{gamma_max}), with an extra parameter $k_B$.  The adopted parameters values to calculate the DM model spectra and LCs are shown in Tab.\ref{Tab:tab_190114c}. The results are shown in Fig. \ref{24fitting190114} in comparison with observational data. 


\begin{table*}
	\caption{Model parameter values for the afterglow of GRB 190114C.}
	\label{Tab:tab_190114c}
	\begin{tabular}{ccccccccc}
		\hline
		 $\alpha_t$ & $\Delta_{\mu}$ & $\epsilon_{B+}$ & $\epsilon_{e}$ & $p$ & $\Gamma_0$ & $E$ $\rm [erg]$ & $n$ $\rm [cm^{-3}]$ & $k_B$\\
		\hline
		$-0.4$ & $10^2$ & $10^{-2}$ & 0.07 & 2.4 & 600 & $2 \times 10^{54}$ & 0.01 & 10\\
		\hline
	\end{tabular}
\centering
\end{table*}

It appears that the model is generally in good agreement with the broad-band data.  
The emission in the 1 keV-100 MeV energy range can be originated from the synchrotron radiation, and the Sub-TeV emission detected by MAGIC is dominated by IC up-scatterings of synchrotron photons. 

Compared with \citet{magic2019a}, which consider the HT model to fit the afterglow data, most parameters adopt in the DM model are similar. But a harder electron spectrum with $p=2.4$ is used here in contrast to $p=2.6$ in \citep{magic2019a}. This is because DM model predicts steeper LC decay than the HT model for the same $p$ value.


The values of parameters inferred from DM model fall within the range of values typically inferred from broadband studies of GRB afterglows, except for the relatively large $E$ value. But the large $E$ is also in line with the Fermi-LAT bright GRBs. These imply that GRB 190114C is not different from a normal long GRB. The Sub-TeV detection of GRB 190114C by MAGIC is simply because it is nearby and with large total ejecta energy. 

We note that by the modeling, the DM model with $\alpha_t \sim -0.4$ is generally consistent with observational data. This value is similar with that derived for the other Fermi-LAT bright GRBs \citep{lemoine13a}.

\section{conclusions}\label{conclusions}\

In this work, we study the afterglow radiation of GRB shocks with DM, and develop a numerical code to calculate the evolution of ED and SED of DM model in a wide range of parameter sets, considering synchrotron and IC radiation with KN effect. 

We compare the DM model with the HT moded, in order to find the characteristic features that the DM model makes. Our numerical results are in good agreement with the anlaytical works by \cite{lemoine13b, lemoine15}. Moreover, our main results for GRB afterglows with DM are:
\begin{enumerate}

\item In the broadband spectrum, the synchrotron radiation in the DM model is similar to that in the HT model with very low magnetic field strength. However, the IC component of the DM model is different from the HT model, because the low energy part of the IC spectral component is produced in the low magnetic field region far downstream of the shock, whereas the high energy part in high magnetic field region close to the shock front. The DM model predicts sharper decaying spectral slope at the high energy end of the IC component than the HT model.

\item The downstream magnetic field structure affects significantly the ED and SED in the afterglow. The faster magnetic field decay, i.e., with smaller decay power-law exponent $\alpha_t$, or with smaller undecaying characteristic scale $\Delta_{\mu}$, the ED and SED deviate more significantly from the standard HT afterglow model, and the synchrotron and IC flux become lower and the IC component moves more toward higher energy. Relatively $\alpha_t$ has larger effect than $\Delta_\mu$ on shaping the ED and SED profiles.


\item The spectrum in the DM model evolves faster than in the HT model, so that the multi-band LCs in the DM model decay faster than in the HT model. The broad-band spectral evolution, in particular the synchrotron component, and the LCs of the DM model can be well consistent with the special HT model with the magnetic field equipartition parameter evolving with time, $\epsilon_B \propto t^{5 \alpha_t /8}$.

\item The DM model can produce a significant IC emission in the TeV range, but due to the strong KN suppression a very strong IC emission with very large Compton parameter, $Y\gg1$, might not happen.
\end{enumerate}

We apply the DM model to explain the multi-band afterglow data of the first detected sub-TeV GRB, GRB 190114C. Due to the steeper LC decay in the DM model, a harder injected spectrum, $p=2.4$, in contrast to $p=2.6$ in the HT model  \citep{magic2019a}, is adopt in the modeling. Moreover, the modeling of the broadband spectrum and multi-band LCs, including the synchrotron and IC components, suggests that a magnetic field decay exponent $\alpha_t \sim -0.4$ can be derived from the data. The future TeV observations of the IC emission from GRBs, e.g., by MAGIC \citep{magic2016a,magic2016b}, HESS \citep{hess2015}, HAWC \citep{hawc2017}, LHAASO \citep{lhaaso2019} and CTA \citep{cta2022,cta2019} will be very helpful to study the microphysics of relativistic shocks.


\appendix
\section{Comparison with Lemoine's works} \label{test}\

\begin{figure}
\vskip -0.0 true cm
\centering
\includegraphics[scale=0.4]{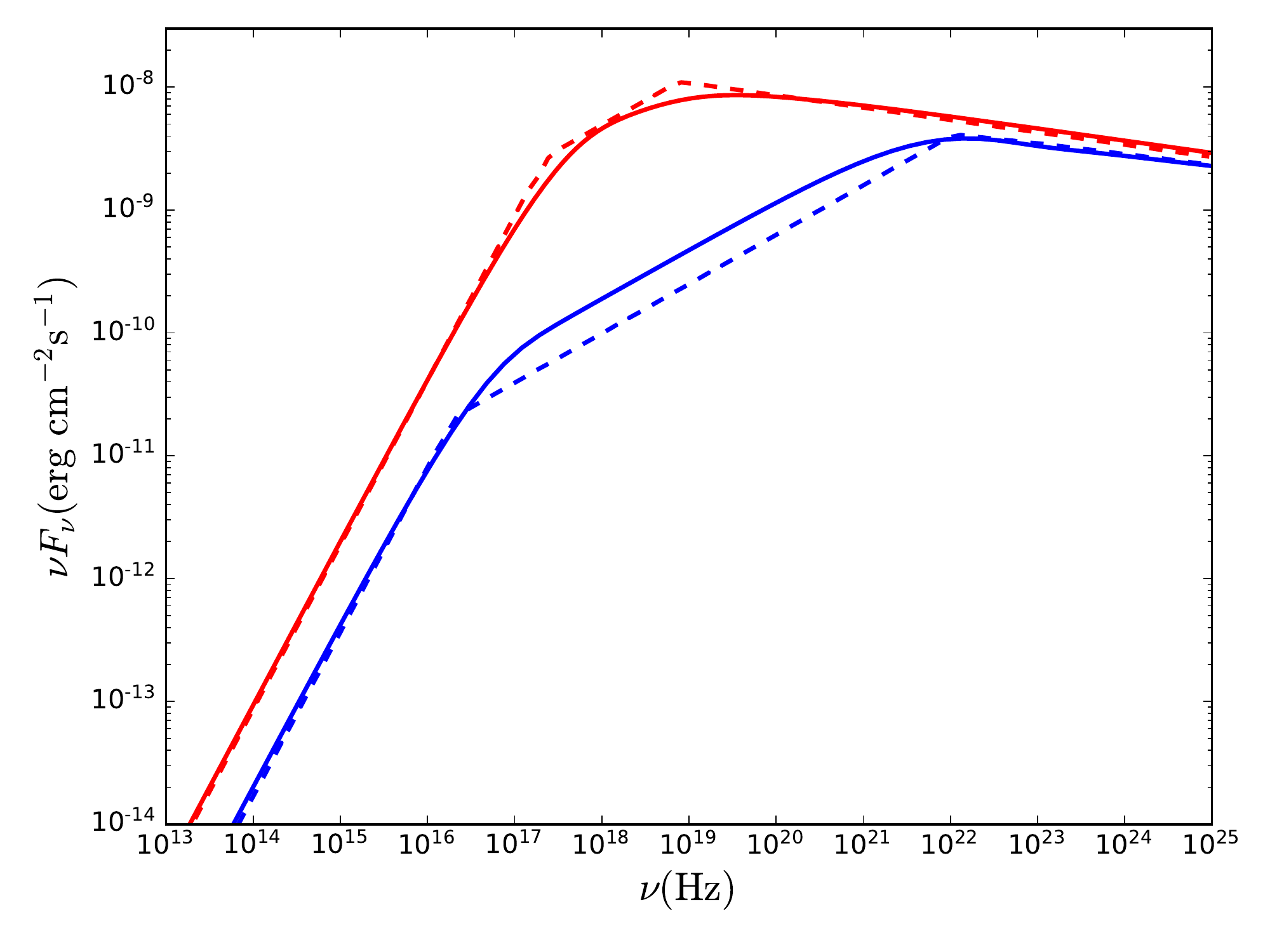}
\includegraphics[scale=0.4]{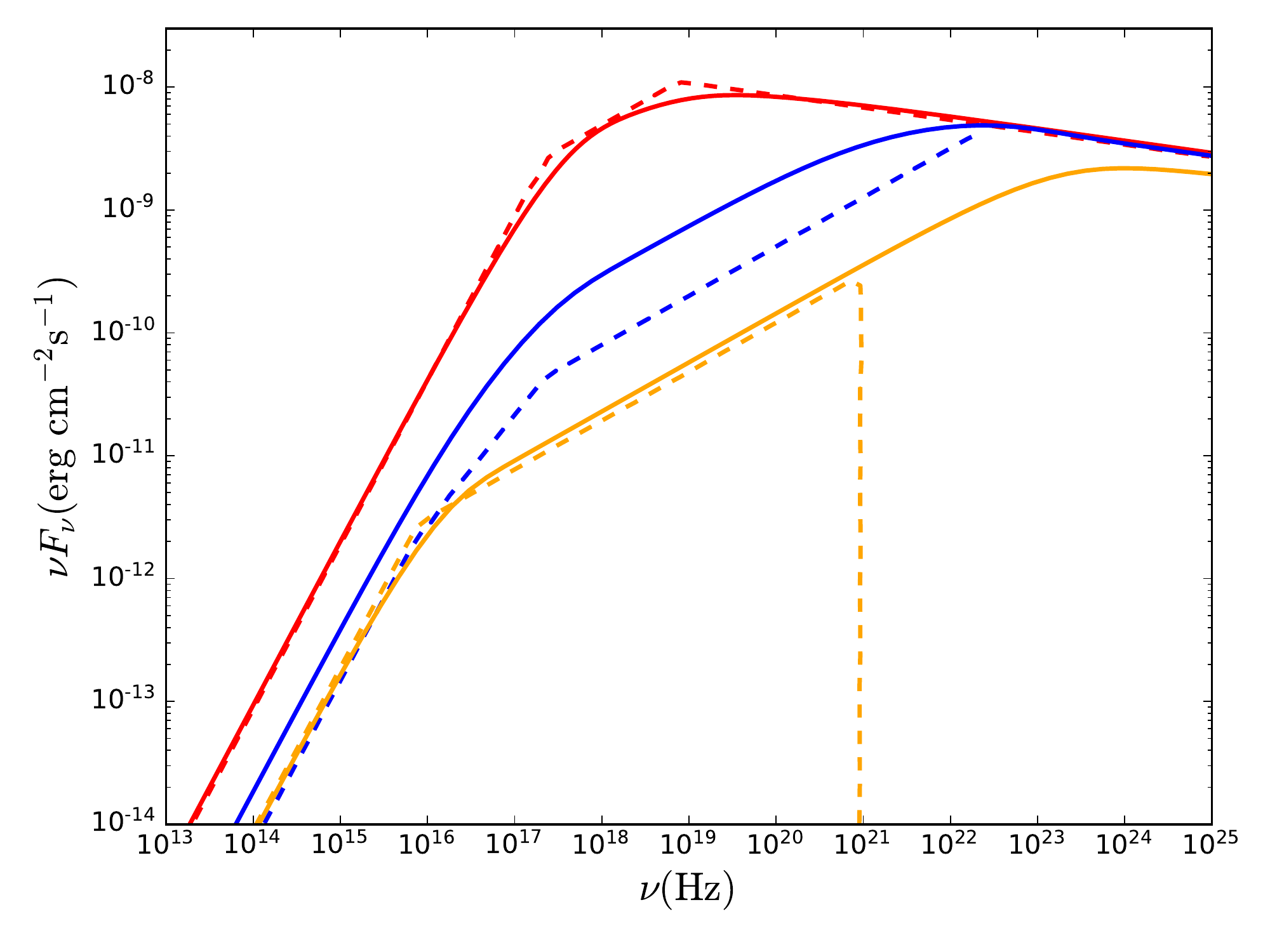}
\includegraphics[scale=0.4]{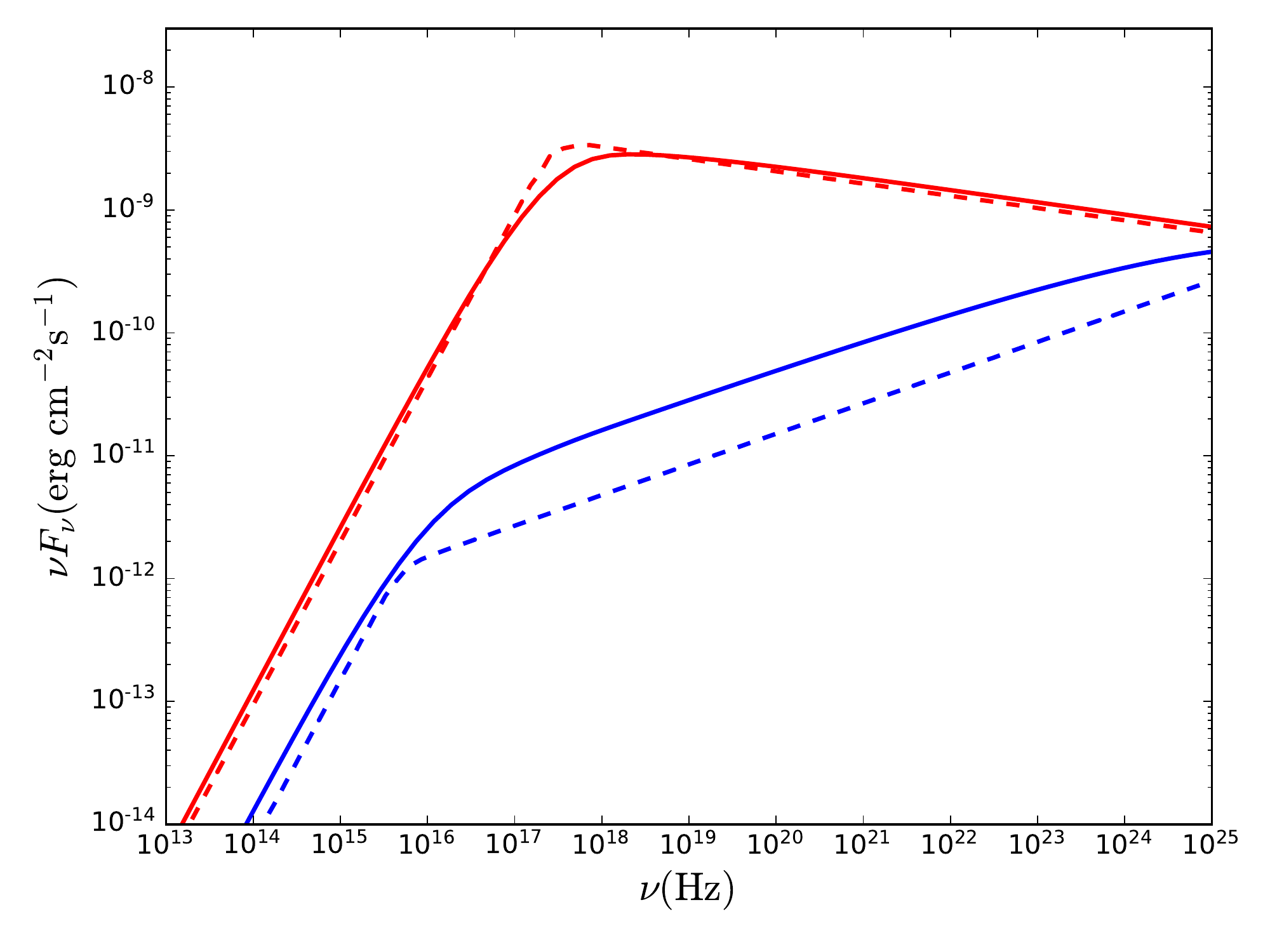}
\caption{Comparison of the synchrotron spectra from numerical simulation in this work (solid lines) and the analytical approach by \citet{lemoine13b} (dashed lines; from Fig1-3 in \citet{lemoine13b}). DM model (blue lines) is compared with HT model (red line) in three typical cases -- Upper panel: gradual decay DM but no IC energy losses ($\alpha_t=-0.5$ and $\Delta_{\mu}=10^2$) versus HT but no IC energy losses ($\epsilon_{B}=10^{-2}$); Middle panel: rapidly decay DM but no IC energy losses ($\alpha_t=-1.8$ and $\Delta_{\mu}=2.7 \times 10^4$) versus HT but no IC energy losses (with $\epsilon_{B}=10^{-2}$ [red line] and $B'=B'_{d}$ [orange line], respectively); Lower panel: DM with strong IC energy losses ($\alpha_t=-0.8$, $\Delta_{\mu}=10^2$, and the Compton parameter immediately behind the shock front $Y_\mu=3$) versus HT with strong IC energy losses ($\epsilon_{B}=10^{-2}$, and constant Compton parameter $Y=3$).}
\label{Fig:compare1}
\end{figure}

\citet{lemoine13b} carried out analytical analysis of the synchrotron spectral signature from relativistic blast waves with DM. Here we compare our results with \citet{lemoine13b} in order to test our numerical code. 

Three cases are considered in \citet{lemoine13b} (fig. 1-3 therein), including the gradual decaying DM ($\alpha_t>-1$) without IC energy losses, rapidly decaying DM ($\alpha_t<-1$) without IC energy losses, and strong IC energy losses. \citet{lemoine13b} expresses the corresponding SEDs and LCs as broken power law functions, and by comparing with HT model, show that the spectro-temporal evolution of the synchrotron spectrum of DM model is changed in contrast with the HT model.  

Here, we adopt the same parameter values as in \citet{lemoine13b}. The calculated synchrotron SEDs are shown in Fig.\ref{Fig:compare1}. One can see that for HT models, the numerical results match well with the analytical ones, whereas for DM model, there are general consistence in the spectral shapes though the flux shows difference up to a factor of 2-3 in the middle spectral segments. It should be noted that in the middle panel, the cutoff in the yellow dashed line is simply because of a high energy cutoff adopted by \citet{lemoine13b} for some reason.

The main flux difference appears in the middle segment corresponding to electrons with $\gamma_{\rm min}'<\gamma_e'<\gamma_c'$, This may be due to the approximation in the analytical approach that the electrons swept up and accelerated by the shock are assumed to be relocated to a region with a distance larger than the fluid reaches in the limited proper time, thus the lower magnetic field leads to lower synchrotorn flux estimated.

\section*{Acknowledgements}\

We thank Martin Lemoine, Katsuaki Asano, Donald C. Warren, Jin-Jun Geng, Xiang-Yu Wang and Yuan-Pei Yang for helpful comments and discussions. YH is supported in part by the National Key R$\&$D Program of China (2021YFC2203100, 2021YFC2203104), the Natural Science Foundation of Anhui Province (2208085QA22), Anhui project (Z010118169) and the Key Research Foundation of Education Ministry of Anhui Province (KJ2020A0008). ZL is supported by the Natural Science Foundation of China (No. 11773003, U1931201) and the China Manned Space Project (CMS-CSST-2021-B11)


\section*{Data Availability}
The data used in this paper were collected from the previous literature. These multi-band data from radio to gamma-ray observations are public for all researchers. 





\bibliographystyle{mnras}
\bibliography{bibliography} 

\begin{thebibliography}{}
\makeatletter
\relax
\def\mn@urlcharsother{\let\do\@makeother \do\$\do\&\do\#\do\^\do\_\do\%\do\~}
\def\mn@doi{\begingroup\mn@urlcharsother \@ifnextchar [ {\mn@doi@}
  {\mn@doi@[]}}
\def\mn@doi@[#1]#2{\def\@tempa{#1}\ifx\@tempa\@empty \href
  {http://dx.doi.org/#2} {doi:#2}\else \href {http://dx.doi.org/#2} {#1}\fi
  \endgroup}
\def\mn@eprint#1#2{\mn@eprint@#1:#2::\@nil}
\def\mn@eprint@arXiv#1{\href {http://arxiv.org/abs/#1} {{\tt arXiv:#1}}}
\def\mn@eprint@dblp#1{\href {http://dblp.uni-trier.de/rec/bibtex/#1.xml}
  {dblp:#1}}
\def\mn@eprint@#1:#2:#3:#4\@nil{\def\@tempa {#1}\def\@tempb {#2}\def\@tempc
  {#3}\ifx \@tempc \@empty \let \@tempc \@tempb \let \@tempb \@tempa \fi \ifx
  \@tempb \@empty \def\@tempb {arXiv}\fi \@ifundefined
  {mn@eprint@\@tempb}{\@tempb:\@tempc}{\expandafter \expandafter \csname
  mn@eprint@\@tempb\endcsname \expandafter{\@tempc}}}

\bibitem[\protect\citeauthoryear{{Abeysekara} et~al.,}{{Abeysekara}
  et~al.}{2017}]{hawc2017}
{Abeysekara} A.~U.,  et~al., 2017, \mn@doi [\apj] {10.3847/1538-4357/aa7555},
  \href {https://ui.adsabs.harvard.edu/abs/2017ApJ...843...39A} {843, 39}

\bibitem[\protect\citeauthoryear{{Achterberg}, {Gallant}, {Kirk}  \&
  {Guthmann}}{{Achterberg} et~al.}{2001}]{ach01}
{Achterberg} A.,  {Gallant} Y.~A.,  {Kirk} J.~G.,   {Guthmann} A.~W.,  2001,
  \mn@doi [\mnras] {10.1046/j.1365-8711.2001.04851.x}, \href
  {https://ui.adsabs.harvard.edu/abs/2001MNRAS.328..393A} {328, 393}

\bibitem[\protect\citeauthoryear{{Ajello} et~al.,}{{Ajello}
  et~al.}{2020}]{ajello2020}
{Ajello} M.,  et~al., 2020, \mn@doi [\apj] {10.3847/1538-4357/ab5b05}, \href
  {https://ui.adsabs.harvard.edu/abs/2020ApJ...890....9A} {890, 9}

\bibitem[\protect\citeauthoryear{{Aleksi{\'c}} et~al.,}{{Aleksi{\'c}}
  et~al.}{2016a}]{magic2016a}
{Aleksi{\'c}} J.,  et~al., 2016a, \mn@doi [Astroparticle Physics]
  {10.1016/j.astropartphys.2015.04.004}, \href
  {https://ui.adsabs.harvard.edu/abs/2016APh....72...61A} {72, 61}

\bibitem[\protect\citeauthoryear{{Aleksi{\'c}} et~al.,}{{Aleksi{\'c}}
  et~al.}{2016b}]{magic2016b}
{Aleksi{\'c}} J.,  et~al., 2016b, \mn@doi [Astroparticle Physics]
  {10.1016/j.astropartphys.2015.02.005}, \href
  {https://ui.adsabs.harvard.edu/abs/2016APh....72...76A} {72, 76}

\bibitem[\protect\citeauthoryear{{Barniol Duran} \& {Kumar}}{{Barniol Duran} \&
  {Kumar}}{2011}]{Barniol2011}
{Barniol Duran} R.,  {Kumar} P.,  2011, \mn@doi [\mnras]
  {10.1111/j.1365-2966.2011.19369.x}, \href
  {https://ui.adsabs.harvard.edu/abs/2011MNRAS.417.1584B} {417, 1584}

\bibitem[\protect\citeauthoryear{{Blandford} \& {McKee}}{{Blandford} \&
  {McKee}}{1976}]{bla76}
{Blandford} R.~D.,  {McKee} C.~F.,  1976, \mn@doi [Physics of Fluids]
  {10.1063/1.861619}, \href
  {https://ui.adsabs.harvard.edu/abs/1976PhFl...19.1130B} {19, 1130}

\bibitem[\protect\citeauthoryear{{Blumenthal} \& {Gould}}{{Blumenthal} \&
  {Gould}}{1970}]{blu70}
{Blumenthal} G.~R.,  {Gould} R.~J.,  1970, \mn@doi [Reviews of Modern Physics]
  {10.1103/RevModPhys.42.237}, \href
  {https://ui.adsabs.harvard.edu/abs/1970RvMP...42..237B} {42, 237}

\bibitem[\protect\citeauthoryear{{Cao} et~al.,}{{Cao}
  et~al.}{2019}]{lhaaso2019}
{Cao} Z.,  et~al., 2019, arXiv e-prints, \href
  {https://ui.adsabs.harvard.edu/abs/2019arXiv190502773C} {p. arXiv:1905.02773}

\bibitem[\protect\citeauthoryear{{Chang} \& {Cooper}}{{Chang} \&
  {Cooper}}{1970}]{Chang70}
{Chang} J.~S.,  {Cooper} G.,  1970, \mn@doi [Journal of Computational Physics]
  {10.1016/0021-9991(70)90001-X}, \href
  {https://ui.adsabs.harvard.edu/abs/1970JCoPh...6....1C} {6, 1}

\bibitem[\protect\citeauthoryear{{Chang}, {Spitkovsky}  \& {Arons}}{{Chang}
  et~al.}{2008}]{chang08}
{Chang} P.,  {Spitkovsky} A.,   {Arons} J.,  2008, \mn@doi [\apj]
  {10.1086/524764}, \href
  {https://ui.adsabs.harvard.edu/abs/2008ApJ...674..378C} {674, 378}

\bibitem[\protect\citeauthoryear{{Cherenkov Telescope Array Consortium}
  et~al.,}{{Cherenkov Telescope Array Consortium} et~al.}{2019}]{cta2019}
{Cherenkov Telescope Array Consortium} et~al., 2019, {Science with the
  Cherenkov Telescope Array}, \mn@doi{10.1142/10986.
}

\bibitem[\protect\citeauthoryear{{Chiaberge} \& {Ghisellini}}{{Chiaberge} \&
  {Ghisellini}}{1999}]{chi99}
{Chiaberge} M.,  {Ghisellini} G.,  1999, \mn@doi [MNRAS]
  {10.1046/j.1365-8711.1999.02538.x}, \href
  {https://ui.adsabs.harvard.edu/abs/1999MNRAS.306..551C} {306, 551}

\bibitem[\protect\citeauthoryear{{Crusius} \& {Schlickeiser}}{{Crusius} \&
  {Schlickeiser}}{1986}]{cru86}
{Crusius} A.,  {Schlickeiser} R.,  1986, Astronomy and Astrophysics, \href
  {https://ui.adsabs.harvard.edu/abs/1986A&A...164L..16C} {164, L16}

\bibitem[\protect\citeauthoryear{{Derishev} \& {Piran}}{{Derishev} \&
  {Piran}}{2019a}]{derishev2007}
{Derishev} E.,  {Piran} T.,  2019a, \mn@doi [ApJL] {10.3847/2041-8213/ab2d8a},
  \href {https://ui.adsabs.harvard.edu/abs/2019ApJ...880L..27D} {880, L27}

\bibitem[\protect\citeauthoryear{{Derishev} \& {Piran}}{{Derishev} \&
  {Piran}}{2019b}]{derishev2019}
{Derishev} E.,  {Piran} T.,  2019b, \mn@doi [ApJL] {10.3847/2041-8213/ab2d8a},
  \href {https://ui.adsabs.harvard.edu/abs/2019ApJ...880L..27D} {880, L27}

\bibitem[\protect\citeauthoryear{{Fan}, {Piran}, {Narayan}  \& {Wei}}{{Fan}
  et~al.}{2008}]{fan08}
{Fan} Y.-Z.,  {Piran} T.,  {Narayan} R.,   {Wei} D.-M.,  2008, \mn@doi [MNRAS]
  {10.1111/j.1365-2966.2007.12765.x}, \href
  {https://ui.adsabs.harvard.edu/abs/2008MNRAS.384.1483F} {384, 1483}

\bibitem[\protect\citeauthoryear{{Fraija}, {Dichiara}, {Pedreira},
  {Galvan-Gamez}, {Becerra}, {Barniol Duran}  \& {Zhang}}{{Fraija}
  et~al.}{2019a}]{fra2019a}
{Fraija} N.,  {Dichiara} S.,  {Pedreira} A.~C. C. d. E.~S.,  {Galvan-Gamez} A.,
   {Becerra} R.~L.,  {Barniol Duran} R.,   {Zhang} B.~B.,  2019a, \mn@doi
  [\apjl] {10.3847/2041-8213/ab2ae4}, \href
  {https://ui.adsabs.harvard.edu/abs/2019ApJ...879L..26F} {879, L26}

\bibitem[\protect\citeauthoryear{{Fraija}, {Barniol Duran}, {Dichiara}  \&
  {Beniamini}}{{Fraija} et~al.}{2019b}]{fra2019b}
{Fraija} N.,  {Barniol Duran} R.,  {Dichiara} S.,   {Beniamini} P.,  2019b,
  \mn@doi [\apj] {10.3847/1538-4357/ab3ec4}, \href
  {https://ui.adsabs.harvard.edu/abs/2019ApJ...883..162F} {883, 162}

\bibitem[\protect\citeauthoryear{{Fukushima}, {To}, {Asano}  \&
  {Fujita}}{{Fukushima} et~al.}{2017}]{fuk17}
{Fukushima} T.,  {To} S.,  {Asano} K.,   {Fujita} Y.,  2017, \mn@doi [ApJ]
  {10.3847/1538-4357/aa7b83}, \href
  {https://ui.adsabs.harvard.edu/abs/2017ApJ...844...92F} {844, 92}

\bibitem[\protect\citeauthoryear{{Geng}, {Huang}, {Wu}, {Zhang}  \&
  {Zong}}{{Geng} et~al.}{2018}]{geng18}
{Geng} J.-J.,  {Huang} Y.-F.,  {Wu} X.-F.,  {Zhang} B.,   {Zong} H.-S.,  2018,
  \mn@doi [ApJS] {10.3847/1538-4365/aa9e84}, \href
  {https://ui.adsabs.harvard.edu/abs/2018ApJS..234....3G} {234, 3}

\bibitem[\protect\citeauthoryear{{Goodman} \& {MacFadyen}}{{Goodman} \&
  {MacFadyen}}{2008}]{goo08}
{Goodman} J.,  {MacFadyen} A.,  2008, \mn@doi [Journal of Fluid Mechanics]
  {10.1017/S0022112008001249}, \href
  {https://ui.adsabs.harvard.edu/abs/2008JFM...604..325G} {604, 325}

\bibitem[\protect\citeauthoryear{{Granot}, {Piran}  \& {Sari}}{{Granot}
  et~al.}{1999}]{gra1999}
{Granot} J.,  {Piran} T.,   {Sari} R.,  1999, \mn@doi [\apj] {10.1086/306884},
  \href {https://ui.adsabs.harvard.edu/abs/1999ApJ...513..679G} {513, 679}

\bibitem[\protect\citeauthoryear{{Gruzinov} \& {Waxman}}{{Gruzinov} \&
  {Waxman}}{1999}]{gru99}
{Gruzinov} A.,  {Waxman} E.,  1999, \mn@doi [ApJ] {10.1086/306720}, \href
  {https://ui.adsabs.harvard.edu/abs/1999ApJ...511..852G} {511, 852}

\bibitem[\protect\citeauthoryear{{He}, {Wu}, {Toma}, {Wang}  \&
  {M{\'e}sz{\'a}ros}}{{He} et~al.}{2011}]{he2011}
{He} H.-N.,  {Wu} X.-F.,  {Toma} K.,  {Wang} X.-Y.,   {M{\'e}sz{\'a}ros} P.,
  2011, \mn@doi [\apj] {10.1088/0004-637X/733/1/22}, \href
  {https://ui.adsabs.harvard.edu/abs/2011ApJ...733...22H} {733, 22}

\bibitem[\protect\citeauthoryear{{Holler} et~al.,}{{Holler}
  et~al.}{2015}]{hess2015}
{Holler} M.,  et~al., 2015, arXiv e-prints, \href
  {https://ui.adsabs.harvard.edu/abs/2015arXiv150902902H} {p. arXiv:1509.02902}

\bibitem[\protect\citeauthoryear{{Keshet}, {Katz}, {Spitkovsky}  \&
  {Waxman}}{{Keshet} et~al.}{2009}]{Keshet09}
{Keshet} U.,  {Katz} B.,  {Spitkovsky} A.,   {Waxman} E.,  2009, \mn@doi [ApJL]
  {10.1088/0004-637X/693/2/L127}, \href
  {https://ui.adsabs.harvard.edu/abs/2009ApJ...693L.127K} {693, L127}

\bibitem[\protect\citeauthoryear{{Kumar} \& {Barniol Duran}}{{Kumar} \&
  {Barniol Duran}}{2009}]{kumar2009}
{Kumar} P.,  {Barniol Duran} R.,  2009, \mn@doi [MNRAS]
  {10.1111/j.1745-3933.2009.00766.x}, \href
  {https://ui.adsabs.harvard.edu/abs/2009MNRAS.400L..75K} {400, L75}

\bibitem[\protect\citeauthoryear{{Kumar} \& {Barniol Duran}}{{Kumar} \&
  {Barniol Duran}}{2010}]{kumar2010}
{Kumar} P.,  {Barniol Duran} R.,  2010, \mn@doi [MNRAS]
  {10.1111/j.1365-2966.2010.17274.x}, \href
  {https://ui.adsabs.harvard.edu/abs/2010MNRAS.409..226K} {409, 226}

\bibitem[\protect\citeauthoryear{{Kumar} \& {Zhang}}{{Kumar} \&
  {Zhang}}{2015}]{kumar2015}
{Kumar} P.,  {Zhang} B.,  2015, \mn@doi [Physics Reports]
  {10.1016/j.physrep.2014.09.008}, \href
  {https://ui.adsabs.harvard.edu/abs/2015PhR...561....1K} {561, 1}

\bibitem[\protect\citeauthoryear{{Laskar} et~al.,}{{Laskar}
  et~al.}{2019}]{laskar2019}
{Laskar} T.,  et~al., 2019, \mn@doi [ApJL] {10.3847/2041-8213/ab2247}, \href
  {https://ui.adsabs.harvard.edu/abs/2019ApJ...878L..26L} {878, L26}

\bibitem[\protect\citeauthoryear{{Lemoine}}{{Lemoine}}{2013}]{lemoine13b}
{Lemoine} M.,  2013, \mn@doi [MNRAS] {10.1093/mnras/sts081}, \href
  {https://ui.adsabs.harvard.edu/abs/2013MNRAS.428..845L} {428, 845}

\bibitem[\protect\citeauthoryear{{Lemoine}}{{Lemoine}}{2015a}]{lemoine15}
{Lemoine} M.,  2015a, \mn@doi [Journal of Plasma Physics]
  {10.1017/S0022377814000920}, \href
  {https://ui.adsabs.harvard.edu/abs/2015JPlPh..81a4501L} {81, 455810101}

\bibitem[\protect\citeauthoryear{{Lemoine}}{{Lemoine}}{2015b}]{lem2015}
{Lemoine} M.,  2015b, \mn@doi [MNRAS] {10.1093/mnras/stv1800}, \href
  {https://ui.adsabs.harvard.edu/abs/2015MNRAS.453.3772L} {453, 3772}

\bibitem[\protect\citeauthoryear{{Lemoine}, {Li}  \& {Wang}}{{Lemoine}
  et~al.}{2013}]{lemoine13a}
{Lemoine} M.,  {Li} Z.,   {Wang} X.-Y.,  2013, \mn@doi [MNRAS]
  {10.1093/mnras/stt1494}, \href
  {https://ui.adsabs.harvard.edu/abs/2013MNRAS.435.3009L} {435, 3009}

\bibitem[\protect\citeauthoryear{{Liu} \& {Wang}}{{Liu} \&
  {Wang}}{2011}]{liu2011}
{Liu} R.-Y.,  {Wang} X.-Y.,  2011, \mn@doi [\apj] {10.1088/0004-637X/730/1/1},
  \href {https://ui.adsabs.harvard.edu/abs/2011ApJ...730....1L} {730, 1}

\bibitem[\protect\citeauthoryear{{L{\'o}pez-Oramas} et~al.,}{{L{\'o}pez-Oramas}
  et~al.}{2022}]{cta2022}
{L{\'o}pez-Oramas} A.,  et~al., 2022, in 37th International Cosmic Ray
  Conference. 12-23 July 2021. Berlin. p.~784 (\mn@eprint {arXiv} {2108.03911})

\bibitem[\protect\citeauthoryear{{MAGIC Collaboration} et~al.,}{{MAGIC
  Collaboration} et~al.}{2019a}]{magic2019b}
{MAGIC Collaboration} et~al., 2019a, \mn@doi [Nature]
  {10.1038/s41586-019-1750-x}, \href
  {https://ui.adsabs.harvard.edu/abs/2019Natur.575..455M} {575, 455}

\bibitem[\protect\citeauthoryear{{MAGIC Collaboration} et~al.,}{{MAGIC
  Collaboration} et~al.}{2019b}]{magic2019a}
{MAGIC Collaboration} et~al., 2019b, \mn@doi [Nature]
  {10.1038/s41586-019-1754-6}, \href
  {https://ui.adsabs.harvard.edu/abs/2019Natur.575..459M} {575, 459}

\bibitem[\protect\citeauthoryear{{Martins}, {Fonseca}, {Silva}  \&
  {Mori}}{{Martins} et~al.}{2009}]{mar09}
{Martins} S.~F.,  {Fonseca} R.~A.,  {Silva} L.~O.,   {Mori} W.~B.,  2009,
  \mn@doi [ApJL] {10.1088/0004-637X/695/2/L189}, \href
  {https://ui.adsabs.harvard.edu/abs/2009ApJ...695L.189M} {695, L189}

\bibitem[\protect\citeauthoryear{{Medvedev} \& {Loeb}}{{Medvedev} \&
  {Loeb}}{1999}]{med99}
{Medvedev} M.~V.,  {Loeb} A.,  1999, \mn@doi [ApJ] {10.1086/308038}, \href
  {https://ui.adsabs.harvard.edu/abs/1999ApJ...526..697M} {526, 697}

\bibitem[\protect\citeauthoryear{{Medvedev}, {Fiore}, {Fonseca}, {Silva}  \&
  {Mori}}{{Medvedev} et~al.}{2005}]{med05}
{Medvedev} M.~V.,  {Fiore} M.,  {Fonseca} R.~A.,  {Silva} L.~O.,   {Mori}
  W.~B.,  2005, \mn@doi [ApJL] {10.1086/427921}, \href
  {https://ui.adsabs.harvard.edu/abs/2005ApJ...618L..75M} {618, L75}

\bibitem[\protect\citeauthoryear{{Medvedev}, {Frederiksen}, {Haugb{\o}lle}  \&
  {Nordlund}}{{Medvedev} et~al.}{2011}]{Medvedev11}
{Medvedev} M.~V.,  {Frederiksen} J.~T.,  {Haugb{\o}lle} T.,   {Nordlund}
  {\r{A}}.,  2011, \mn@doi [ApJ] {10.1088/0004-637X/737/2/55}, \href
  {https://ui.adsabs.harvard.edu/abs/2011ApJ...737...55M} {737, 55}

\bibitem[\protect\citeauthoryear{{Misra} et~al.,}{{Misra}
  et~al.}{2021}]{misra2019}
{Misra} K.,  et~al., 2021, \mn@doi [\mnras] {10.1093/mnras/stab1050}, \href
  {https://ui.adsabs.harvard.edu/abs/2021MNRAS.504.5685M} {504, 5685}

\bibitem[\protect\citeauthoryear{{Nakar}, {Ando}  \& {Sari}}{{Nakar}
  et~al.}{2009}]{nakar2009}
{Nakar} E.,  {Ando} S.,   {Sari} R.,  2009, \mn@doi [\apj]
  {10.1088/0004-637X/703/1/675}, \href
  {https://ui.adsabs.harvard.edu/abs/2009ApJ...703..675N} {703, 675}

\bibitem[\protect\citeauthoryear{{Panaitescu}}{{Panaitescu}}{2005}]{pan05}
{Panaitescu} A.,  2005, \mn@doi [MNRAS] {10.1111/j.1365-2966.2005.09532.x},
  \href {https://ui.adsabs.harvard.edu/abs/2005MNRAS.363.1409P} {363, 1409}

\bibitem[\protect\citeauthoryear{{Panaitescu} \& {Kumar}}{{Panaitescu} \&
  {Kumar}}{2000}]{pan00}
{Panaitescu} A.,  {Kumar} P.,  2000, \mn@doi [\apj] {10.1086/317090}, \href
  {https://ui.adsabs.harvard.edu/abs/2000ApJ...543...66P} {543, 66}

\bibitem[\protect\citeauthoryear{{Panaitescu} \& {Kumar}}{{Panaitescu} \&
  {Kumar}}{2002}]{pan02}
{Panaitescu} A.,  {Kumar} P.,  2002, \mn@doi [ApJ] {10.1086/340094}, \href
  {https://ui.adsabs.harvard.edu/abs/2002ApJ...571..779P} {571, 779}

\bibitem[\protect\citeauthoryear{{Pennanen}, {Vurm}  \& {Poutanen}}{{Pennanen}
  et~al.}{2014}]{pen14}
{Pennanen} T.,  {Vurm} I.,   {Poutanen} J.,  2014, \mn@doi [Astronomy and
  Astrophysics] {10.1051/0004-6361/201322520}, \href
  {https://ui.adsabs.harvard.edu/abs/2014A&A...564A..77P} {564, A77}

\bibitem[\protect\citeauthoryear{{Petropoulou} \& {Mastichiadis}}{{Petropoulou}
  \& {Mastichiadis}}{2009}]{pet09}
{Petropoulou} M.,  {Mastichiadis} A.,  2009, \mn@doi [Astronomy and
  Astrophysics] {10.1051/0004-6361/200912970}, \href
  {https://ui.adsabs.harvard.edu/abs/2009A&A...507..599P} {507, 599}

\bibitem[\protect\citeauthoryear{{Piran}}{{Piran}}{2005}]{Piran05}
{Piran} T.,  2005, in {de Gouveia dal Pino} E.~M.,  {Lugones} G.,   {Lazarian}
  A.,  eds,  American Institute of Physics Conference Series Vol. 784, Magnetic
  Fields in the Universe: From Laboratory and Stars to Primordial Structures..
  pp 164--174 (\mn@eprint {arXiv} {astro-ph/0503060}),
  \mn@doi{10.1063/1.2077181}

\bibitem[\protect\citeauthoryear{{Ravasio} et~al.,}{{Ravasio}
  et~al.}{2019}]{rav2019}
{Ravasio} M.~E.,  et~al., 2019, \mn@doi [Astronomy and Astrophysics]
  {10.1051/0004-6361/201935214}, \href
  {https://ui.adsabs.harvard.edu/abs/2019A&A...626A..12R} {626, A12}

\bibitem[\protect\citeauthoryear{{Rhoads}}{{Rhoads}}{1999}]{rhoads99}
{Rhoads} J.~E.,  1999, \mn@doi [ApJ] {10.1086/307907}, \href
  {https://ui.adsabs.harvard.edu/abs/1999ApJ...525..737R} {525, 737}

\bibitem[\protect\citeauthoryear{{Rossi} \& {Rees}}{{Rossi} \&
  {Rees}}{2003}]{rossi2003}
{Rossi} E.,  {Rees} M.~J.,  2003, \mn@doi [MNRAS]
  {10.1046/j.1365-8711.2003.06242.x}, \href
  {https://ui.adsabs.harvard.edu/abs/2003MNRAS.339..881R} {339, 881}

\bibitem[\protect\citeauthoryear{{Rybicki} \& {Lightman}}{{Rybicki} \&
  {Lightman}}{1979}]{ryb79}
{Rybicki} G.~B.,  {Lightman} A.~P.,  1979, {Radiative processes in
  astrophysics}

\bibitem[\protect\citeauthoryear{{Santana}, {Barniol Duran}  \&
  {Kumar}}{{Santana} et~al.}{2014}]{Santana14}
{Santana} R.,  {Barniol Duran} R.,   {Kumar} P.,  2014, \mn@doi [ApJ]
  {10.1088/0004-637X/785/1/29}, \href
  {https://ui.adsabs.harvard.edu/abs/2014ApJ...785...29S} {785, 29}

\bibitem[\protect\citeauthoryear{{Sari} \& {Esin}}{{Sari} \&
  {Esin}}{2001}]{sari01}
{Sari} R.,  {Esin} A.~A.,  2001, \mn@doi [ApJ] {10.1086/319003}, \href
  {https://ui.adsabs.harvard.edu/abs/2001ApJ...548..787S} {548, 787}

\bibitem[\protect\citeauthoryear{{Sari}, {Piran}  \& {Narayan}}{{Sari}
  et~al.}{1998}]{sari98}
{Sari} R.,  {Piran} T.,   {Narayan} R.,  1998, \mn@doi [ApJL] {10.1086/311269},
  \href {https://ui.adsabs.harvard.edu/abs/1998ApJ...497L..17S} {497, L17}

\bibitem[\protect\citeauthoryear{{Sironi} \& {Goodman}}{{Sironi} \&
  {Goodman}}{2007}]{sir07}
{Sironi} L.,  {Goodman} J.,  2007, \mn@doi [ApJ] {10.1086/523636}, \href
  {https://ui.adsabs.harvard.edu/abs/2007ApJ...671.1858S} {671, 1858}

\bibitem[\protect\citeauthoryear{{Sironi} \& {Spitkovsky}}{{Sironi} \&
  {Spitkovsky}}{2009}]{sir09}
{Sironi} L.,  {Spitkovsky} A.,  2009, \mn@doi [ApJL]
  {10.1088/0004-637X/707/1/L92}, \href
  {https://ui.adsabs.harvard.edu/abs/2009ApJ...707L..92S} {707, L92}

\bibitem[\protect\citeauthoryear{{Sironi} \& {Spitkovsky}}{{Sironi} \&
  {Spitkovsky}}{2011}]{sir11}
{Sironi} L.,  {Spitkovsky} A.,  2011, \mn@doi [ApJ]
  {10.1088/0004-637X/741/1/39}, \href
  {https://ui.adsabs.harvard.edu/abs/2011ApJ...741...39S} {741, 39}

\bibitem[\protect\citeauthoryear{{Spitkovsky}}{{Spitkovsky}}{2008}]{spi08}
{Spitkovsky} A.,  2008, \mn@doi [ApJL] {10.1086/590248}, \href
  {https://ui.adsabs.harvard.edu/abs/2008ApJ...682L...5S} {682, L5}

\bibitem[\protect\citeauthoryear{{Wang}, {Li}, {Moradi}  \& {Ruffini}}{{Wang}
  et~al.}{2019}]{wangxy2019}
{Wang} Y.,  {Li} L.,  {Moradi} R.,   {Ruffini} R.,  2019, arXiv e-prints, \href
  {https://ui.adsabs.harvard.edu/abs/2019arXiv190107505W} {p. arXiv:1901.07505}

\bibitem[\protect\citeauthoryear{{Waxman}}{{Waxman}}{1997}]{wax1997}
{Waxman} E.,  1997, \mn@doi [\apjl] {10.1086/311057}, \href
  {https://ui.adsabs.harvard.edu/abs/1997ApJ...491L..19W} {491, L19}

\bibitem[\protect\citeauthoryear{{Weibel}}{{Weibel}}{1959}]{wei59}
{Weibel} E.~S.,  1959, \mn@doi [PRL] {10.1103/PhysRevLett.2.83}, \href
  {https://ui.adsabs.harvard.edu/abs/1959PhRvL...2...83W} {2, 83}

\bibitem[\protect\citeauthoryear{{Wijers} \& {Galama}}{{Wijers} \&
  {Galama}}{1999}]{wij99}
{Wijers} R.~A.~M.~J.,  {Galama} T.~J.,  1999, \mn@doi [ApJ] {10.1086/307705},
  \href {https://ui.adsabs.harvard.edu/abs/1999ApJ...523..177W} {523, 177}

\bibitem[\protect\citeauthoryear{{Zhang}}{{Zhang}}{2019}]{zhang2019}
{Zhang} B.,  2019, \mn@doi [Nature] {10.1038/d41586-019-03503-6}, \href
  {https://ui.adsabs.harvard.edu/abs/2019Natur.575..448Z} {575, 448}

\bibitem[\protect\citeauthoryear{{Zhang}, {MacFadyen}  \& {Wang}}{{Zhang}
  et~al.}{2009}]{zha09}
{Zhang} W.,  {MacFadyen} A.,   {Wang} P.,  2009, \mn@doi [ApJL]
  {10.1088/0004-637X/692/1/L40}, \href
  {https://ui.adsabs.harvard.edu/abs/2009ApJ...692L..40Z} {692, L40}

\bibitem[\protect\citeauthoryear{{Zhang}, {Christie}, {Petropoulou},
  {Rueda-Becerril}  \& {Giannios}}{{Zhang} et~al.}{2020}]{zhanghao2019}
{Zhang} H.,  {Christie} I.~M.,  {Petropoulou} M.,  {Rueda-Becerril} J.~M.,
  {Giannios} D.,  2020, \mn@doi [\mnras] {10.1093/mnras/staa1583}, \href
  {https://ui.adsabs.harvard.edu/abs/2020MNRAS.496..974Z} {496, 974}

\bibitem[\protect\citeauthoryear{{Zirakashvili} \& {Aharonian}}{{Zirakashvili}
  \& {Aharonian}}{2007}]{zir07}
{Zirakashvili} V.~N.,  {Aharonian} F.,  2007, \mn@doi [Astronomy and
  Astrophysics] {10.1051/0004-6361:20066494}, \href
  {https://ui.adsabs.harvard.edu/abs/2007A&A...465..695Z} {465, 695}

\makeatother
\end{thebibliography}




\bsp	
\label{lastpage}

\end{document}